\documentclass{aa}

\usepackage{graphicx}
\usepackage{amsmath}
\usepackage{subcaption}
\usepackage{mwe}
\usepackage{booktabs}
\usepackage{tabularx}
\usepackage{natbib}
\usepackage{xcolor}
\bibpunct{(}{)}{;}{a}{}{,}
\setcitestyle{citesep={;}}
\usepackage{gensymb}
\usepackage{graphicx}
\usepackage{txfonts}
\usepackage{siunitx}
\newcommand{\mockalph}[1]{}

\begin{document}

   \title{X-ray emission from a 
   rapidly accreting narrow-line Seyfert 1 galaxy 
   at z=6.56}

   \author{J. Wolf\,\thanks{jwolf@mpe.mpg.de}
          \inst{1,2},
          K. Nandra
          \inst{1},
          M. Salvato
          \inst{1}, 
          J. Buchner
          \inst{1}, 
          M. Onoue
          \inst{3,4}, 
          T. Liu
          \inst{1}, 
          R. Arcodia
          \inst{1}, 
          A. Merloni
          \inst{1}, 
          S. Ciroi
          \inst{5}, 
          \newline
          F. Di Mille
          \inst{6},   
          V. Burwitz,
          \inst{1}, 
          M. Brusa
          \inst{7,8}, 
          R. Ishimoto
          \inst{9}, 
          N. Kashikawa
          \inst{9}, 
          Y. Matsuoka
          \inst{10}, 
          T. Urrutia
          \inst{11}
          S. G. H. Waddell
          \inst{1}
          }
          
    \titlerunning{eROSITA high-z NLS1}
    \authorrunning{Wolf et al.}

   \institute{Max-Planck-Institut f\"{u}r extraterrestrische Physik, Gie\ss enbachstra\ss e 1, 85748 Garching, Germany
         \and
         Exzellenzcluster ORIGINS, Boltzmannstr. 2, D-85748 Garching, Germany 
         \and Kavli Institute for Astronomy and Astrophysics, Peking University, Beijing 100871, China
         \and Kavli Institute for the Physics and Mathematics of the Universe (Kavli IPMU, WPI), The University of Tokyo, Chiba 277-8583, Japan
         \and Dipartimento di Fisica e Astronomia dell’ Universit\`{a} di Padova ,Vicolo dell’Osservatorio 3, I-35122 Padova, Italy
         \and Las Campanas Observatory - Carnegie Institution for Science, Colina el Pino, Casilla 601, La Serena, Chile
         \and Dipartimento di Fisica e Astronomia "Augusto Righi", Alma Mater Studiorum Università di Bologna, via Gobetti 93/2, 40129 Bologna, Italy
         \and INAF - Osservatorio di Astrofisica e Scienza dello Spazio di Bologna, via Gobetti 93/3, 40129 Bologna, Italy
         \and Department of Astronomy, Graduate School of Science, The University of Tokyo, 7-3-1 Hongo, Bunkyo, Tokyo 113-0033, Japan
         \and Research Center for Space and Cosmic Evolution, Ehime University, Matsuyama, Ehime 790-8577, Japan
         \and Leibniz-Institut für Astrophysik Potsdam (AIP). An der Sternwarte 16. 14482 Potsdam, Germany}

   \date{Received August 08, 2022; accepted November 02, 2022}

  \abstract
   {   The space density of X-ray-luminous, blindly selected active galactic nuclei (AGN) traces the population of rapidly accreting super-massive black holes through cosmic time. It is encoded in the X-ray luminosity function, whose bright end remains poorly constrained in the first billion years after the Big Bang as X-ray surveys have thus far lacked the required cosmological volume. With the eROSITA Final Equatorial-Depth Survey (eFEDS), the largest contiguous and homogeneous X-ray survey to date, X-ray AGN population studies can now be extended to new regions of the luminosity--redshift space ($L_{\mathrm{2-10 \, keV}} >10^{45} \, \mathrm{erg \, s^{-1}}$ and $z>6$).
  }
  {The current study aims at identifying luminous quasars at $z>5.7$ among X-ray-selected sources in the eFEDS field in order to place a lower limit on black hole accretion well into the epoch of re-ionisation. A secondary goal is the characterisation of the physical properties of these extreme coronal emitters at high redshifts.
  }
   {Cross-matching eFEDS catalogue sources to optical counterparts from the DESI Legacy Imaging Surveys, we confirm the low significance detection with eROSITA of a previously known, optically faint $z=6.56$ quasar from the Subaru High-z Exploration of Low-luminosity Quasars (SHELLQs) survey. We obtained a pointed follow-up observation of the source with the \textit{Chandra} X-ray telescope in order to confirm the low-significance eROSITA detection. Using new near-infrared spectroscopy, we derived the physical properties of the super-massive black hole. Finally, we used this detection to infer a lower limit on the black hole accretion density rate at $z>6$.
   }
   {The \textit{Chandra} observation confirms the eFEDS source as the most distant blind X-ray detection to date. The derived X-ray luminosity is high with respect to the rest-frame optical emission of the quasar. With a narrow Mg{\sc ii} line, low derived black hole mass, and high Eddington ratio, as well as its steep photon index, the source shows properties that are similar to local narrow-line Seyfert 1 galaxies, which are thought to be powered by young super-massive black holes. In combination with a previous high-redshift quasar detection in the field, we show that quasars with $L_{2-10 \, \mathrm{keV}} >10^{45} \, \mathrm{erg \, s^{-1}}$ dominate accretion onto super-massive black holes at $z\sim 6$.}
   {}

   \keywords{quasars: individual --
                Galaxies: high-redshift --
                X-rays: galaxies
               }

   \maketitle

\section{Introduction}

Over the last 20 years, quasars have been discovered at ever-increasing redshifts and well into the epoch of re-ionisation \citep[e.g.][]{fan01,willott09,mortlock12,wu15,jiang16,matsuoka18,banados16,wang21}. These objects signpost accretion onto super-massive black holes (SMBHs) through cosmic time. The mere existence of $>10^9 \, M_\odot$ black holes in the first gigayear of the Universe ($z>5.7$, \citealt{onoue19,yang21}) challenges SMBH seeding and growth models, requiring sustained Eddington-limited or even super-Eddington accretion \citep{volonteri05}. The nature of the seeds themselves is still being investigated \citep[for reviews, see][]{volonteri10,haiman13,johnson16,latif16,volonteri21}. While the bulk of active galaxies discovered at $z>5.7$ host black holes with masses of 1 to 10 billion solar masses, there must be a population of less massive ($~10^8 \, M_\odot$), super-Eddington-accreting SMBHs caught in an earlier evolutionary state. Some of these less massive, strongly accreting black holes have been found at the centre of high-redshift quasars that display optical properties similar to local narrow-line Seyfert 1 galaxies \citep[NLS1s; e.g.][]{koptelova17,koptelova19,onoue19}. NLS1s are a special class of active galaxies that are defined by their narrow $\mathrm{H}\beta$ emission lines ($\mathrm{FWHM}<2000 \, \mathrm{km \, s^{-1}}$) and the weakness of their $ [\mathrm{O}\textsc{iii}]$ narrow-line emission relative to $\mathrm{H}\beta$, $\mathrm{[O\textsc{iii}]}/\mathrm{H}\beta<3$ \citep{osterbrock83,goodrich89}. They show strong Fe{\sc ii} emission \citep{osterbrock85}, typically host SMBHs with lower black hole masses ($M_{\mathrm{BH}}<10^8 \, M_\odot$), and accrete at a significant fraction of their Eddington limit \citep[$10-100\%$; e.g.][]{pounds95,grupe10,rakshit17,waddell20}, as expected from young and strongly accreting black holes. Large amplitude, short timescale flaring behaviour in the UV continuum has been observed for this class of source \citep{collier01}. 
Rapid, high amplitude variability 
 is
 also seen at shorter wavelengths,
 in X-rays \citep[e.g.][]{turner2001, romano02}. NLS1s usually have steeper X-ray spectra (i.e. larger power-law photon indices). than typical broad-line Seyfert 1 galaxies \citep{nandra94,boller96}. 

 Beyond the end of the epoch of re-ionisation, hard X-ray photons unhindered by dust and gas are collectable by sensitive soft X-ray telescopes at observer-frame energies, $\sim 2 \, \mathrm{keV}$. To date, $\sim 50$ quasars at $z>5.7$ have been observed in X-rays, mostly via pointed observations with \textit{Chandra} and \textit{XMM-Newton} \citep{brandt02,vignali03,nanni17,vito19} following their discovery with optical telescopes. Recently, \citet{barlow_hall22} reported the blind detection of a quasar spectroscopically confirmed at $z=6.31$ in the Extragalactic Serendipitous Swift Survey (ExSeSS). The first unbiased, blind X-ray detections of quasars at $z>5.7$ in the performance verification fields and all-sky maps of the extended ROentgen Survey with an Imaging Telescope Array (eROSITA; \citealt{predehl21,sunyaev21}) were reported by \citet{medvedev20} and \citet{wolf21}. \citet{khorunzev21} present the discovery with eROSITA of the most X-ray-luminous quasar beyond $z \gtrsim 5.5$ with $\log L_{\rm 2-10 \, \mathrm{keV}}=3\times 10^{46} \, \mathrm{erg\, s^{-1}}$. In addition to being among the most X-ray-luminous quasars at the end of re-ionisation, all of the eROSITA-detected high-redshift quasars are radio detected and radio loud (according to the radio-loudness definition $R=f_{\rm \nu,5\, GHz}/f_{\rm \nu,4400\, \AA}>10$ defined by \citealt{kellerman89}).
 However, the $z=5.81$ quasar detected in the eROSITA Final Equatorial Depth Survey \citep[eFEDS;][]{brunner21} does not show evidence of any jet contribution in its X-ray output, making it a secure probe of coronal activity and hence black hole accretion history \citep{wolf21}. The direct X-ray selection of this spectroscopically confirmed quasar in a contiguous field of uniform exposure imposes constraints on the X-ray luminosity function (XLF) just after the epoch of re-ionisation ($z\sim 6$). \citet{wolf21} show that an exponential decline at high redshift cannot be excluded. However, models that show a shallower slope at the bright end of the XLF are preferred in the probed high-redshift bin.    

Here we present the eROSITA X-ray detection of a second high-redshift quasar in the eFEDS field: the $z=6.56$ quasar J0921+0007, initially discovered in a dedicated survey based on the Hyper Suprime Cam (HSC) Subaru Strategic Program \citep[SSP;][]{aihara22}: the Subaru High-z Exploration of Low-luminosity Quasars \citep[SHELLQs;][]{matsuoka18}. The source is optically faint but X-ray bright. Its optical and near-infrared (NIR) spectral properties potentially make it a high-redshift NLS1. We present the eROSITA detection of this optically faint source with a 21 ks \textit{Chandra} ACIS-S follow-up observation and derive its X-ray properties in Sect. \ref{sec:det}. 
We derive the black hole mass of the source and Eddington ratio with a new Ks-band spectrum that covers the Mg{\sc ii} region in Sect. \ref{sec:mw}. We connect the global optical and X-ray emission in Sect. \ref{sec:x_opt} and show that J0921+0007 is X-ray loud.  It is the highest-redshift blindly detected X-ray source to date, and its detection further supports a flattening of the XLF beyond the break luminosity, $L_*$. We present its contribution to the global accretion density in Sect. \ref{sec:demographics}.

We have assumed a standard $\mathrm{\Lambda}$ cold dark matter cosmology with parameters from \citet{planck18}. Throughout this work, uncertainties are quoted at the 68\% confidence level (1$\sigma$).

 \section{HSC J092120.56+000722.9: An X-ray-luminous quasar}
\label{sec:det}
\subsection{eROSITA detection}

The eFEDS was executed during the eROSITA performance verification phase and covers approximately 140 deg$^2$ to a nominal exposure of 2.2 ks.
In eFEDS, sources were detected in the $0.2-2.3 \, \rm keV$ band with the
\texttt{erbox} task of the eROSITA Science Analysis Software System \citep[eSASS;][]{brunner21}. A detection likelihood threshold $\mathrm{DET\_LIKE} \geq 6$ was applied. A supplementary catalogue\footnote{https://erosita.mpe.mpg.de/edr/eROSITAObservations/Catalogues/} of 4774 eROSITA/eFEDS sources detected just below this threshold ($ 5<\mathrm{DET\_LIKE} < 6$) was also made available. It is expected to contain a high fraction of spurious sources, but it also gives access to interesting faint objects. 
Using simulations, \citet{brunner21} show that reducing the source detection likelihood threshold from $\mathrm{DET\_LIKE}=6$ to $\mathrm{DET\_LIKE}=5$  results in an increase in the number of detections of the simulated point-sources. They report an increase 1\%  (94 \% instead of 93 \%) of detected simulated point-sources brighter than $F_{0.5-2\, \mathrm{keV}}=1\times 10^{-14} \, \mathrm{erg\, s^{-1} cm^{-2}}$ and 4\% (63 \% instead of 59 \%) of sources brighter than $F_{0.5-2\, \mathrm{keV}}=4\times 10^{-15} \, \mathrm{erg\, s^{-1} cm^{-2}}$. Similarly, \citet{liu21}, performing a standard single-band (0.2-2.3 keV) detection run on a simulated eFEDS field, show that reducing the $\rm DET\_LIKE$ threshold from 6 to 5 results in an overall increase in completeness from $\sim 79 \%$ to $\sim 82 \%$ and an increase in the spurious fraction from $\sim 7 \%$ to $\sim 12 \%$. For the supplementary sample, we identified optical counterparts in the eighth data release of the Dark Energy Spectroscopic Instrument Legacy Imaging Surveys \citep[hereafter LS8;][]{dey19} using the Bayesian cross-matching algorithm \texttt{NWAY} \citep{salvato18}. It computes posterior cross-match probabilities, accounting for the surface densities of the matched catalogues, astrometric distances, and uncertainties, as well as independent photometric information. For the latter, we applied the random-forest-generated `photometric prior' presented by \citet{salvato21}, a model trained on an independent 3XMM point-source population and its optical LS8 properties.

 The resulting best optical counterpart solutions were positionally cross-matched to a complete list of spectroscopically confirmed $z>5.5$ quasars in the eFEDS footprint compiled from literature \citep[31 sources, ][]{fan01,venemans15,matsuoka16,matsuoka18,matsuoka18b,matsuoka22} within $1''$. This exercise returned one match. The eFEDS source J092120.6+000725.9, hereafter J0921+0007, has a LS8 match within $3.11''$ of its centroid, which coincides spatially with the spectroscopically confirmed SHELLQs quasar J0921+0007 \citep[$z=6.56$,][]{matsuoka18}  within $0.08''$. The spectroscopic redshift of this quasar was securely measured by \citet{matsuoka18}, using the Ly$\alpha$ emission line. It was subsequently confirmed by Onoue et al. (in prep.) using Mg{\sc ii} and \citet{yang22} who found $z= 6.5646\pm 0.0003$ using [C{\sc ii}]. There were 22 possible LS8 counterpart candidates within $30''$ of the X-ray source J0921+0007. 
 The individual probability for the chosen LS8 counterpart of being the correct one amongst the candidates is unequivocally high $\texttt{p\_i}=0.82$ (other candidates have $\texttt{p\_i}<0.13$).
A summary of the \texttt{NWAY} match is presented in Table \ref{crossmatch}.

\begin{table}[]
\begin{tabular}{@{}lcl@{}}
\toprule
eROSITA ID          & -            & 22224  \\ 
eROSITA Name          & -            & eFEDS J0921+0007  \\ 
$\rm RA_{eFEDS}$           & {[}deg{]}    & $140.3361$                 \\
$\rm DEC_{eFEDS}$          & {[}deg{]}    & $0.1237$                   \\
$\rm \sigma_{RADEC,eFEDS}$ & {[}arcsec{]} & $3.87$                      \\
$\rm DET\_LIKE$ (0.2-2.3 keV)            & -            & $5.04$                    \\
Net Counts (0.2-2.3 keV)         & - & $10.5 \pm 4.7$                      \\
LS8 objID/brickID          & -            & 5281/330929               \\
$\rm Sep. eRO/LS8$           & {[}arcsec{]} & $2.77$                       \\
QSO ID                     & -            & HSC J0921+0007 \\
QSO Redshift               & -            & 6.56                    \\
$\rm Sep. QSO/LS8$         & {[}arcsec{]} & \textless{}0.1           \\ \bottomrule
\end{tabular}
    \caption{Basic source and counterpart information. The coordinates of the eFEDS source are equatorial, with $\rm \sigma_{RADEC, \,eFEDS}$ being the $1\sigma$ X-ray positional uncertainty. The net counts and errors are obtained via photon-mode PSF fitting \citep{brunner21}. Sep. eRO/LS8 measures the separation between the centroid of the eFEDS X-ray source and the position of the LS8 counterpart. Sep. QSO/LS8 corresponds to the separation between the quasar optical position and the LS8 counterpart. }
    \label{crossmatch}
\end{table}

\begin{figure}[]
\includegraphics[width=.4\textwidth]{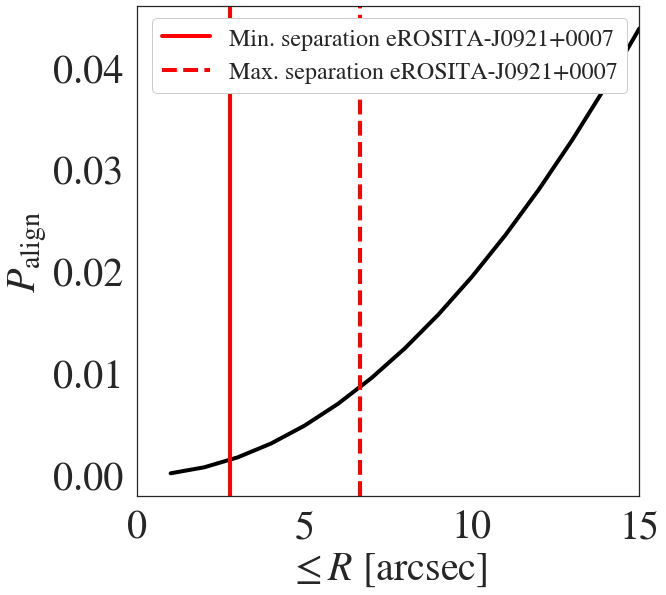}

\caption{Probability of finding any of the 31 spectroscopically confirmed quasars in the eFEDS footprint within a distance $R$ of a spurious X-ray source. The solid (dashed) red line shows the minimum (maximum)  distance between the eROSITA source and J0921+0007, accounting for the positional uncertainty of the X-ray source. }
\label{fig:chance}
\centering
\end{figure}

While the detection likelihood is at a low level where a large spurious fraction is expected, the alignment with a high-redshift quasar strengthens the detection. At an X-ray detection likelihood of 5, 12\% of sources in the eFEDS field are expected to be spurious \citep{liu21}. We estimated the probability of a chance alignment of any spectroscopically confirmed high-redshift quasar in the eFEDS footprint with a spurious detection (i.e. a background fluctuation detected as catalogue source). For this we first observe that, in the eFEDS footprint excluding the borders with lower exposure, higher background and stronger vignetting \citep[90\% of the total area; see][]{liu22} there are 29482 sources detected in the 0.2-2.3 keV band above the detection likelihood of the quasar $\mathrm{DET\_LIKE}>5.04$. Among these sources, 3277.6 are expected to be spurious from simulations. We thus obtained the probability of a chance alignment as a function of the separating distance $R$ of a quasar and a spurious source as: $31\times 3277.6 \pi R^2/\mathrm{area}_{90}$. Here $\mathrm{area}_{90}$ is the area of the `$90 \%$' region: 1640219392 $\mathrm{arcsec}^2$. The evolution of the probability of a spurious chance alignment with radius is shown in Fig. \ref{fig:chance}. Accounting for the eFEDS bi-variate positional error ($\sigma=\mathrm{RADEC\_ERR} / \sqrt{2}=3.87''$), within the maximum separation between J0921+0007 and the eROSITA source, the probability of a chance alignment of the quasar with a spurious source is less than $1\%$.

In order to confirm the eROSITA detection, we obtained a 21ks \textit{Chandra} follow-up observation (GTO proposal, cycle 22, ObsID 24738) pointed at the optical position of the quasar.

\subsection{Confirmation with a Chandra pointed observation}

\begin{figure*}[]
\includegraphics[width=1.\textwidth]{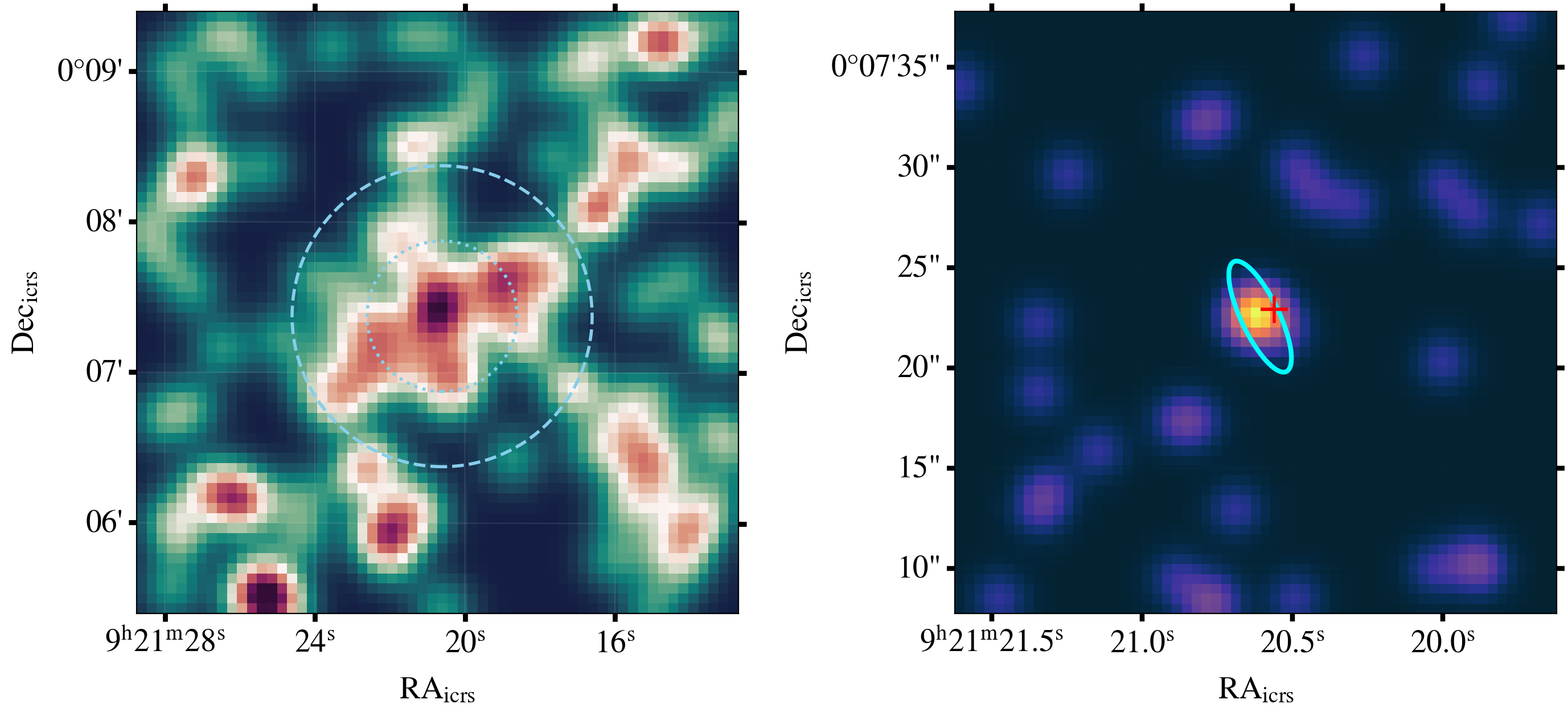}

\caption{X-ray image cutouts in the region of J0921+0007. \textit{Left: }$4' \times 4' $ eROSITA/eFEDS image in the $0.2-2.3 \, \rm keV$ band, smoothed with a Gaussian kernel. The image is centred on the optical position of the quasar J0921+0007. The concentric circles have radii of $30''$ (dotted) and $60''$ (dashed). \textit{Right: }$30'' \times 30'' $ high-resolution \textit{Chandra} image in the energy range 0.5 - 7 keV. The ellipse shows a \textit{wavdetect}-detected source at the optical position of the quasar (marked by a red cross). }
\label{fig:image}
\centering
\end{figure*}

On October 26, 2021, the quasar was observed with \textit{Chandra} ACIS-S with a total exposure time of 21.47 ks (PI:Predehl, Observer: Wolf). On the 0.5 - 7 keV band \textit{Chandra} image, sources were detected with the mexican-hat wavelet algorithm \textit{wavdetect} from the \texttt{CIAO} software package. 
The default detection parameters for the pixel radii  (\texttt{scales}) as well as the significance thresholds (\texttt{sighthresh} and \texttt{bkgsigthresh}) have been used (respectively 2 and 4 pixels, $10^{-6}$ and 0.001). We confirm the
significant detection of an X-ray source whose centroid lies within $0.78''$ of the optical coordinates of the quasar. The positional counts-weighted variances in pixels are $\mathrm{X\_ERR}=0.45$ and $\mathrm{Y\_ERR}=0.28$ ($\mathrm{RA\_ERR}=0.22''$ and $\mathrm{DEC\_ERR}=0.14''$), for a point-spread function (PSF) size of $0.46''$. 

No other source was detected within $30''$ of the optical quasar coordinates in the 0.5 - 7 keV band. A high-resolution broadband image and the \textit{wavdetect} $3\sigma$ elliptical source detection region are presented in the right panel of Fig. \ref{fig:image}.
In parallel to this automated detection procedure, we performed forced photometry at the quasar position on the 0.5-7 keV \textit{Chandra} images. We extracted source counts in a circular region of $2''$ radius centred on the coordinates of the quasar. Background counts were extracted in a ring with inner and outer radii of $4''$ and $20''$. As shown in the right panel of Fig. \ref{fig:image}, no other bright source is present in the background region. We computed the binomial no-source probability \citep[e.g.][]{weisskopf07,vito19} as

\begin{equation}
    P_B(i\geq s) =  \sum_{i=s}^{s+b} \frac{(s+b)!}{i!(i-s-b)!} \left( \frac{1}{1+r} \right)^i\left(\frac{2+r}{1+r}\right)^{s+b-i}  
.\end{equation}Here $s$ and $b$ are counts in the source and background region, while r is given by the ratio of areas of these two extraction regions. In the 0.5-7 keV band, we extracted 7 counts in the source region and 52 counts in the background region.  We obtained $P_{\rm B,0.5-7 \, \mathrm{keV}} \sim 3\times10^{-6}$; in other words, the source detection is highly significant. 
The binomial no source probabilities in the narrower energy bands 0.5 - 1.2 keV and 1.2 - 2 keV are $P_{\rm B, \, 0.5 - 1.2 \, keV} \sim 3\times 10^{-4}$ and $P_{\rm B, \, 1.2 - 2 \, keV} \sim 6\times 10^{-4}$. 

\subsection{X-ray properties}
\label{sec:xray_prop}

 The forced PSF-fitting photometry results in the supplementary eFEDS catalogue impose that J0921+0007 was significantly detected in the 0.5-1 keV band ($\rm DET\_LIKE=6.2$). This band samples the rest-frame hard X-ray emission ($\lambda(\mathrm{0.5 -1 \, keV})\sim 3.8 - 7.6 \rm \, keV$). 
 In this band, there are $4.74 \pm 2.74$ source net counts (as measured from the rate). The background at the source position is $2.42$ counts/$\rm arcmin^2$. The measured count rate is $r_{0.5-1\, \mathrm{keV}} = (3.86 \pm 2.23) \times 10^{-3} cts/s$. We converted the count-rate to an observed frame  
 flux of $F_{0.5-1 \, keV}= (2.8 \pm 1.6) \times 10^{-15} \, \mathrm{erg \, cm^{-2}\, s^{-1}}$ assuming a power law with photon index $\Gamma=2$ and Galactic foreground absorption $N_\mathrm{H}= 2.65\times 10^{20} \, \mathrm{cm^{-2}}$ from  \citet{hi4pi16}\footnote{At this stage the choice of the spectral shape is arbitrary. The aim here is a consistent comparison of the eFEDS and \textit{Chandra} data.}. This corresponds to a rest-frame $2-10 \, \mathrm{keV}$ luminosity of $L_{\rm 2-10 \,keV}= (2.96 \pm 1.71) \times 10^{45} \rm erg\, s^{-1}$. 
 We also computed count rates and fluxes from the \textit{Chandra} follow-up data using the \texttt{srcflux} script of the \texttt{CIAO} software package. We manually selected source and background regions centred at the optical coordinates of the quasar. The circular source region has a radius of $2''$, while the annulus describing the background region has radii ($4'',13''$). Counts were extracted in the 0.5-2 keV band. For the PSF model, we opted for the \texttt{arfcorr} method. 
 We obtained the net count rate: $r_{0.5-2 \, \rm keV}= (2.8^{+2.5}_{-1.6})\times 10^{-4} {\mathrm{cts/s}}$. Once again assuming a nominal absorbed power law with $\rm \Gamma = 2$ and $N_\mathrm{H}= 2.65 \times 10^{20} \, \mathrm{cm^{-2}}$, we obtained the absorption-corrected energy flux 
 $F_{0.5-2.0 \, \mathrm{keV}}= (3.5^{+3.1}_{-2.0}) \times 10^{-15} \, \mathrm{erg \, cm^{-2}\, s^{-1}}$ and the 2-10 keV luminosity $ L_{2-10 \, \rm keV} = (2.1^{+1.9}_{-1.2}) \times 10^{45}  \rm erg\, s^{-1}$. The luminosity derived from the \textit{Chandra} data is consistent with the eROSITA results within the 1$\sigma$ measurement uncertainties. The large error bars are driven by the low-count statistics. Within these uncertainties, no flux variability is detected between the eROSITA observations and \textit{Chandra} (i.e. $\Delta t \sim 95$ days) in the quasars rest frame.

 \begin{table}[]
\begin{tabular}{@{}ccccc@{}}
\toprule
{\color[HTML]{000000} Obs.} & {\color[HTML]{000000} Net rate} & {\color[HTML]{000000} Flux abs.} & {\color[HTML]{000000} $L_{\rm 2-10 \, keV}$} & {\color[HTML]{000000} $\Gamma$}  \\ \midrule
{\color[HTML]{000000} } & {\color[HTML]{000000} $10^{-4}$ cts/s} & {\color[HTML]{000000} $10^{-15} \, \mathrm{erg \, /cm^2/s}$} & {\color[HTML]{000000} $10^{45} \, \mathrm{erg/s}$} & {\color[HTML]{000000} -} \\ \midrule
{\color[HTML]{000000} \textit{\begin{tabular}[c]{@{}c@{}}eROSITA\\ 0.5-1 keV\end{tabular}}} & {\color[HTML]{000000} $38.6 \pm 22.3$} & {\color[HTML]{000000} $2.84 \pm 1.64$} &  {\color[HTML]{000000} $2.96 \pm 1.71$} & {\color[HTML]{000000} $2$} \\
{\color[HTML]{000000} \textit{\begin{tabular}[c]{@{}c@{}}Chandra\\ 0.5-2 keV\end{tabular}}} & {\color[HTML]{000000} $2.79^{+2.47}_{-1.56}$} & {\color[HTML]{000000} $3.26^{+2.87}_{-1.83}$} & {\color[HTML]{000000} $2.10^{+1.86}_{-1.17}$} & {\color[HTML]{000000} $2$} \\
{\color[HTML]{000000} \begin{tabular}[c]{@{}c@{}}\textit{Chandra}\\ \texttt{BXA}\end{tabular}} & {\color[HTML]{000000} -} & {\color[HTML]{000000} $4.43^{+2.78}_{-1.75}$} & {\color[HTML]{000000} $3.72^{+3.97}_{-1.89}$} & {\color[HTML]{000000} $3.2^{+0.7}_{-0.6}$} \\\bottomrule
\end{tabular}
\caption{X-ray properties of J0921+0007. The two first lines present the photometry derived from the eROSITA and \textit{Chandra} observations of J0921+0007 assuming a fixed spectral model ($\Gamma=2$). The flux and luminosities derived from the spectral fit to the \textit{Chandra} data are labelled `\textit{Chandra} \texttt{BXA'}.}
\label{table:xphotometry}
\end{table}

\begin{figure}[]
\includegraphics[width=1.0\columnwidth]{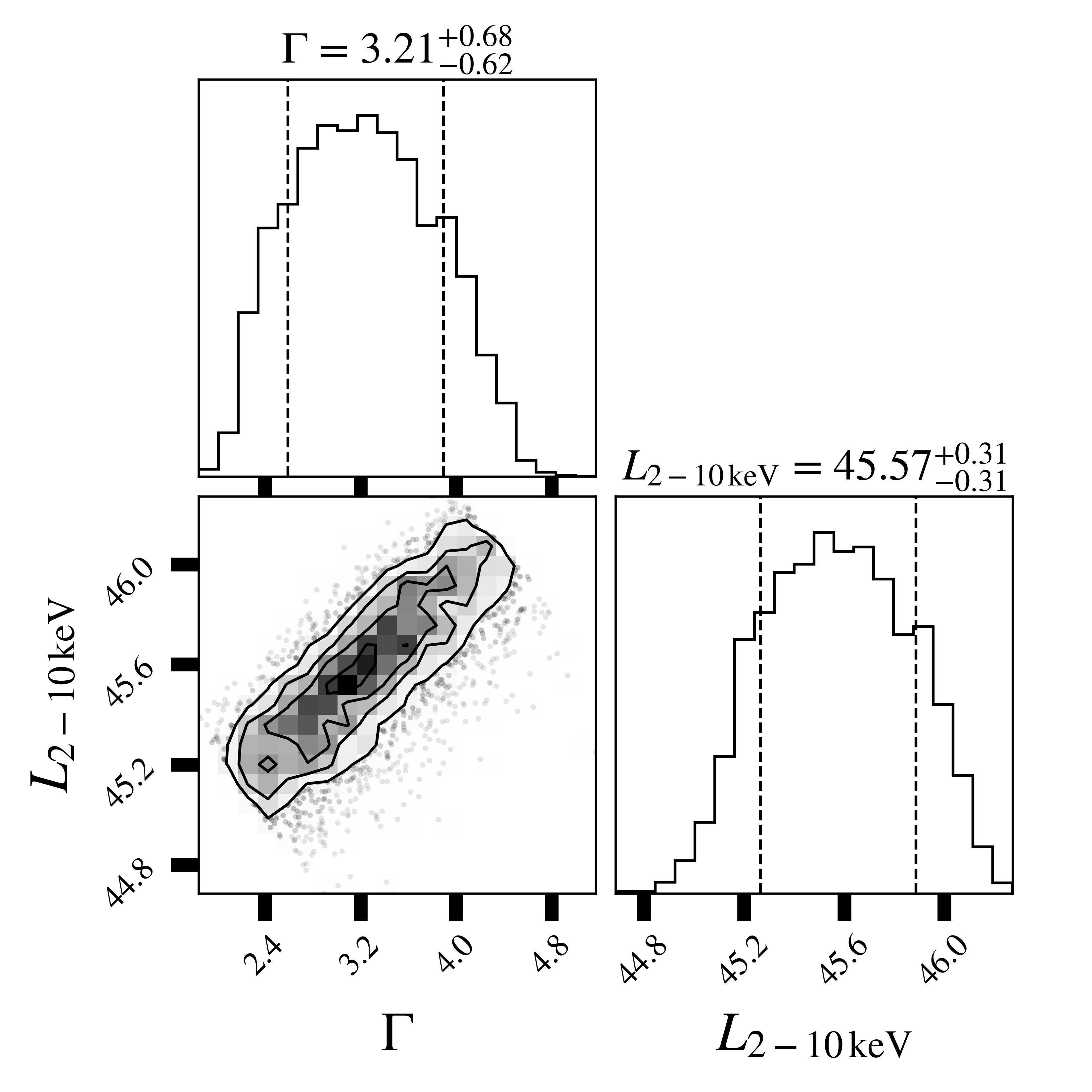}

\caption{Marginal posterior distributions of the photon index, $\Gamma$, and the hard X-ray luminosity from the \texttt{BXA} fit to the \textit{Chandra} spectrum.}
\label{fig:corner}
\centering
\end{figure}

 In addition, we obtained a tentative spectral fit of the counts extracted from the manually defined source and background regions. We used the Bayesian analysis software \texttt{BXA} \citep{buchner14} coupled to the X-ray fitting library \texttt{XSPEC} \citep{arnaud96} and the nested sampling algorithm \texttt{ultranest} \citep{buchner21}. We restricted the fitting region to the 0.1-8.5 keV band. Given the low number of counts, we adopt an absorbed, redshifted power-law model: \textit{tbabs*zpowerlw}, with photoelectric absorption fixed to the Galactic absorbing column density $N_{\rm H}$.
 The normalisation and the photon index $\rm \Gamma$ were allowed to vary in the fit, assuming uninformative, wide priors ($\Gamma=1.5-5$). The redshift of \textit{zpowerlw} was fixed to the redshift of the quasar, $z=6.56$. We used the C-statistic for the spectral analysis \citep{cash79}.
 The marginal posterior distributions of the photon index and $L_{\rm 2-10 \, keV}$ are presented in Fig. \ref{fig:corner}. We obtained a 
 steep power-law photon index $\Gamma=3.2^{+0.7}_{-0.6}$. Even assuming no intrinsic absorption, the photon index is higher than typical quasars in this redshift regime \citep[$\Gamma=2.20$,][]{vito19}. However, we cannot exclude that the steepness of the photon index is due to the presence of a high-energy cutoff, redshifted into the observed \textit{Chandra} waveband. The high energy cut-off corresponds to the temperature of the corona, beyond which electrons are no longer able to give energy to the incident photons. Typically this cut-off is two to three times larger than the coronal electron temperature \citep{petrucci01}, and is often of the order of a few hundred keV, but has been measured as low as a few tens of keV \citep[e.g.][]{kara17}, which is within the observed-frame for this source. Another possible explanation for this steep soft spectrum is the presence of a soft-excess \citep{arnaud85,boller96,magdziarz98,gierlinski04,ross05,crummy06,walton13}. We derived absorbed fluxes and luminosities, accounting for the posterior samples of $\Gamma$ and the power-law normalisation (see Table \ref{table:xphotometry}). We adopted  $\mathrm{\Gamma}=3.2$ for the spectral shape of J0921+0007 throughout this work.

\begin{figure}[]
\includegraphics[width=.5\textwidth]{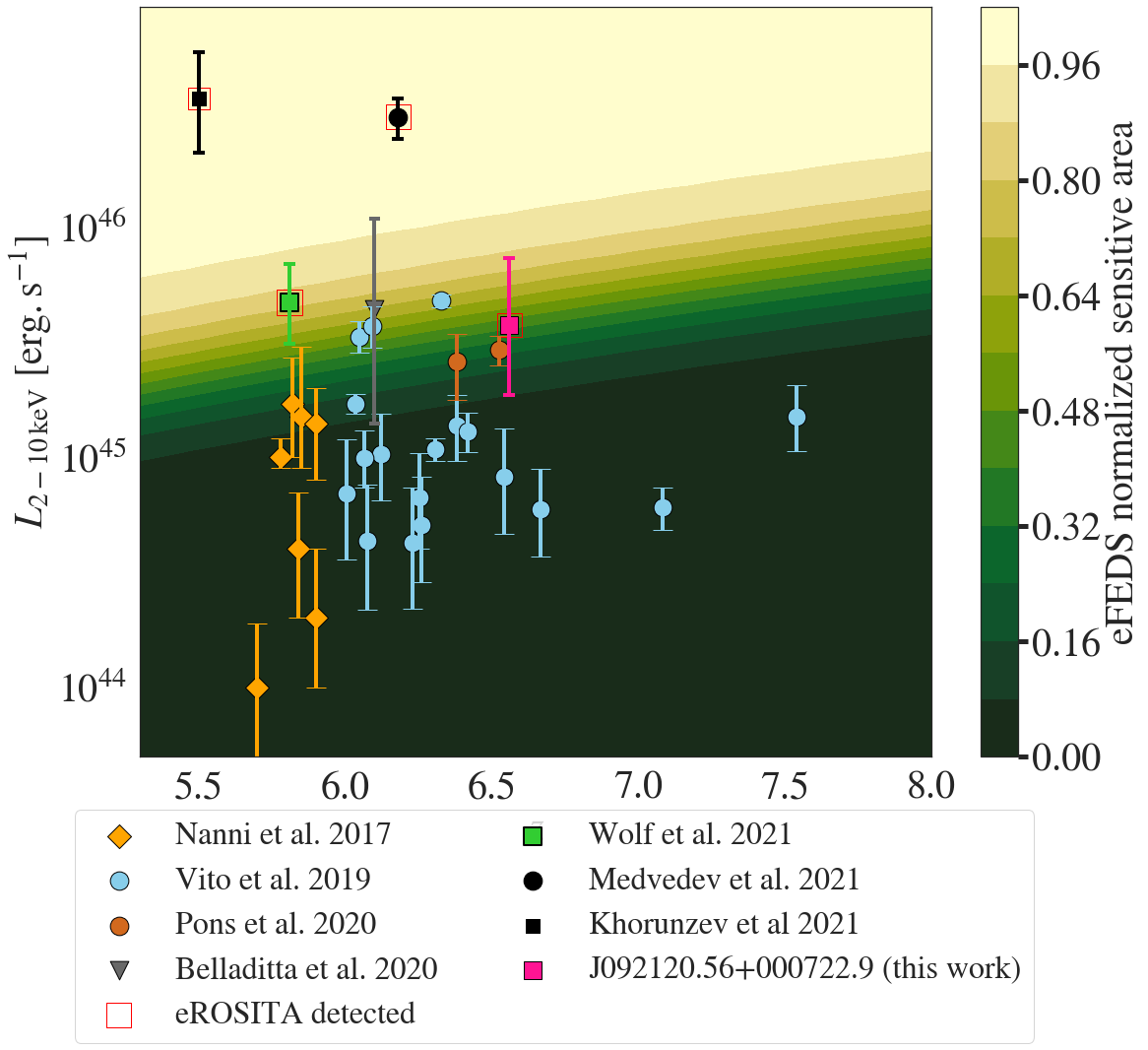}

\caption{Hard X-ray luminosities of X-ray-detected high-redshift quasars. All eROSITA-detected quasars are marked by a red square. The normalised area of eFEDS that is sensitive to sources modelled by a fiducial absorbed power law ($\rm \Gamma = 3$ and $N_\mathrm{H}= 3\times 10^{20} \, \mathrm{cm^{-2}}$) is represented by the background colour map. J0921+0007, shown here as a pink square, lies at the detection limit of the eFEDS survey, as expected from the low eROSITA detection likelihood.}
\label{fig:lx_plot}
\centering
\end{figure}

 Table \ref{table:xphotometry} summarises the X-ray photometry derived from the eROSITA and \textit{Chandra} data. Fig. \ref{fig:lx_plot} presents the 2-10 keV luminosity-redshift plane for an up-to-date sample of X-ray-detected quasars \citep{nanni17,vito19,pons20,belladitta20,medvedev21,khorunzev21,wolf21}. It also displays the eFEDS normalised sensitive area. The area sensitivity curves are obtained with the eSASS task \texttt{apetool}. 
 \texttt{apetool} returns the sensitive area in square degrees as a function of source count-rates.
 We converted count-rates to $L_{\rm 2-10 \, \mathrm{keV}}$ assuming an absorbed power-law model with $\rm \Gamma = 3$ and $N_\mathrm{H}= 3\times 10^{20} \, \mathrm{cm^{-2}}$.  The luminosity of J0921+0007 derived from the spectral analysis is shown. The quasar lies at the sensitivity limit of the survey.

\section{Physical properties and active galactic nucleus type from a $Ks$-band spectrum}
\label{sec:mw}
J0921+0007 was discovered by \citet{matsuoka18} in the dedicated HSC high-redshift quasar survey SHELLQs. It was selected as quasar candidate based on its red $i-z$ colour and retained by a Bayesian selection method detailed by \citet{matsuoka16}. The photometry of this quasar resembles that of Galactic brown dwarfs and could only be disentangled from stellar contaminants due to an unusually strong $\rm Ly\alpha + NV$ complex, which steepens the $z-Y$ slope \citep{matsuoka18}. Indeed, J0921+0007 possesses the second most luminous Lyman $\alpha$ line of the entire SHELLQs $z>5.8$ quasar sample with $\mathrm{log}\, (L_{\rm Ly\alpha}/\mathrm{(erg \, s^{-1})})=45\pm 0.01$ \citep{matsuoka16,matsuoka18}. \citet{matsuoka18} further report a relatively moderate line width $ \mathrm{FWHM_{Ly\alpha}}= \rm 1400 \pm 100 \, km \, s^{-1}$. We present a new NIR observation of this source, derive its black hole properties and determine its AGN type.

\subsection{Black hole mass and accretion rate}
\label{sec:nir_sec}

We estimated the black hole mass from infrared spectroscopy. The Mg{\sc ii} $\lambda$2798 emission line, which can be used as a virial black hole mass estimator \citep{vestergaard09} is redshifted outside the optical discovery spectrum, which ends at 1.02um (1330 $\AA$ in the rest frame). Thus, we obtained NIR spectroscopy of J0921+0007.
 We note that this spectrum was obtained prior to the eROSITA detection of the source.

The $Ks$-band spectrum of J0921+0007 was obtained by MOIRCS \citep{ichikawa06,suzuji08}, a Cassegrain instrument mounted on the Subaru Telescope, on April 22, 2019. The observation was performed in the multi-object spectroscopy mode for secure target acquisition. The VPH-K grism \citep{ebizuka11} was used to cover 2.0 -- 2.4 $\mu$m at a spectral resolution of $R=1700$ for a 0.8'' slit width.
J0921+0007 was observed for 72 minutes with mean airmass 1.1 and $K$-band seeing size 0.8 arcsecond. More details of the observations and data analysis will be presented in Onoue et al. (in prep.).

The raw data were reduced and 1D-extracted in the standard manner based on the software system Image Reduction and Analysis Facility (IRAF). To correct for the telluric absorption an A0-type star was observed just before the exposures of J0921+0007. The telluric-corrected 1D spectrum was then scaled to the $Ks$-band magnitude of J0921+0007 ($=20.691 \pm 0.052$ AB mag) that was obtained by the same run with MOIRCS. Observations in the $Ks$ band are preferable since it covers Mg{\sc ii}, the emission line of interest. This 10 minutes imaging observation enables us to flux-calibrate the observed $K$-band spectrum accurately without being affected by potential variability of the quasar. The spectrum was scaled to correct for the Galactic extinction in the K band. 

\begin{figure}[]
\includegraphics[width=.5\textwidth]{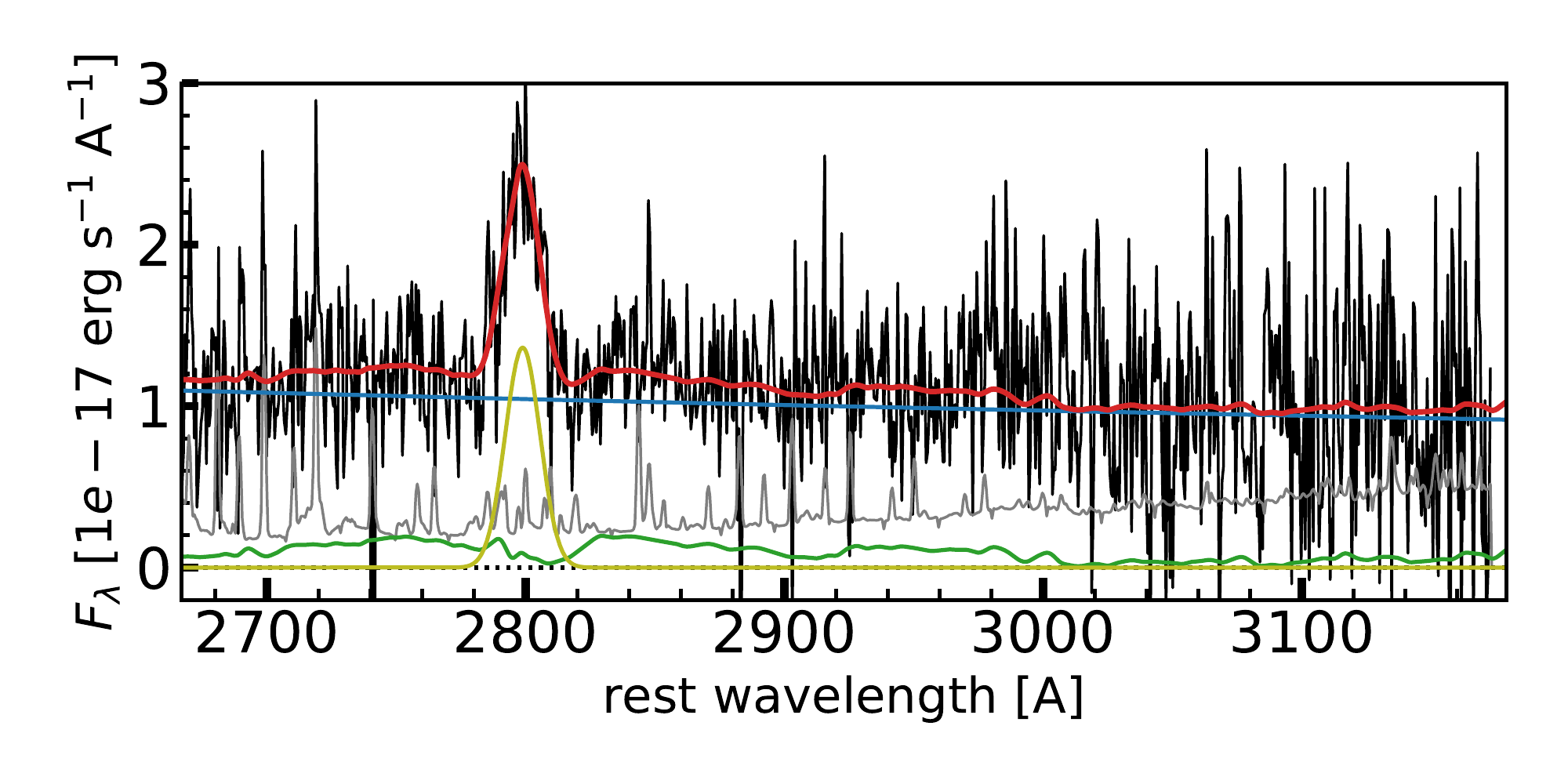}

\caption{$Ks$-band MOIRCS spectrum of J0921+0007. The combined fit to the continuum and Mg{\sc ii} line is shown by the solid red curve. The decomposition in continuum (blue), the Fe{\sc ii} pseudo continuum (green), and the Gaussian line fit (yellow) are also shown.}
\label{fig:nirspec}
\centering
\end{figure}

Figure \ref{fig:nirspec} shows the obtained $Ks$-band spectrum, where a strong Mg{\sc ii} emission line is clearly detected. We model the spectrum with three components: power-law continuum, Fe{\sc ii} pseudo-continuum, and Mg{\sc ii}. Since the MOIRCS spectrum covers a narrow wavelength range, the continuum power-law slope ($\alpha_\lambda=-1.052$; $F_\lambda\propto\lambda^{\alpha_\lambda}$) was estimated by the photometric colour of the optical HSC $y$ band and MOIRCS $Ks$ band. The monochromatic luminosity at rest frame $3000 \, \AA$ $L_{3000} = (4.8 \pm 0.2) \times 10^{42}$ erg s$^{-1}$ s$^{-1}$ $\AA^{-1}$ was derived by the scaled power-law continuum model with $\alpha_\lambda=-1.021$. The 3000 $\AA$ luminosity was then converted to the bolometric luminosity $L_\mathrm{bol}= (7.4 \pm 0.3) \times 10^{46}$ erg s$^{-1}$ assuming a bolometric factor of 5.5 \citep{richards06}. For Fe{\sc ii} emission lines, the empirical template of a local narrow-line Seyfert galaxy, 1 Zw 1 \citep{tsuzuki06} was convolved with a Gaussian kernel and fitted to the observed continuum together with the power-law continuum. A single Gaussian profile was fitted to the residual to measure the Mg{\sc ii} line shape. The derived Mg{\sc ii} redshift of $6.5634^{+0.0013}_{-0.0012}$ is consistent with the Ly$\alpha$ redshift reported by \citet[][$z=6.56$]{matsuoka18}. The redshift measurement is also consistent with the recent [C{\sc ii}] redshift reported by \citet{yang21} for this object ($z=6.5646 \pm 0.0003$).

The BH mass and Eddington ratio were derived based on the Mg{\sc ii} single-epoch method \citep{vestergaard09}. From the Mg{\sc ii} line full width half maximum (FWHM $=1699^{+99}_{-110}$ km s$^{-1}$) and the $3000 \, \AA$  luminosity, 
we measured the virial black hole mass $M_\mathrm{BH} = (2.48^{+0.31}_{-0.29}) \times 10^8 M_\odot$ and an Eddington ratio $\lambda=L_{\rm bol}/L_{\rm Edd}=L_{\rm bol}/(4\pi\, c \, G \, M_{\rm BH} m_\mathrm{p}/\sigma_\mathrm{T})=2.29^{+0.30}_{-0.29}$. J0921+0007 shows a high Eddington ratio, as expected from its steep photon index. 

We note that J0921+0007 was also observed with Gemini/GNIRS by \citet{yang21}. Their continuum and Mg{\sc ii} measurements are mostly consistent with ours, while they report a slightly fainter 3000 angstrom luminosity ($L_{3000} = (3.9 \pm 0.4) \times 10^{42}$ erg s$^{-1}$). This difference is likely attributed to the different absolute flux calibration between the GNIRS and MOIRCS spectra. 
For the calibration of their spectrum, \citet{yang21} used a $J$-band magnitude measurement with a relatively large error ($J=21.21\pm0.28$) for their calibration.

\subsection{NLS1 classification}

 Following  \citet{osterbrock83} and \citet{goodrich89}, a quasar is required to show a narrow H$\beta$ line ($\mathrm{FWHM_{\mathrm{H}\beta}}< 2000 \, \mathrm{km \, s^{ -1}}$) and a small narrow-line  to broad-line flux ratio ($\rm [O\textsc{iii}]/H\beta<3$) in order to be classified as NLS1s.
 At $z> 6$, however, $\rm H\beta$ is shifted out to mid-infrared wavelengths, making the direct classification of high-z quasars as NLS1s by this definition impossible. Using the correlation between Mg{\sc ii} and $\rm H\beta$ widths in a sample of Sloan Digital Sky Survey DR14 quasars, \citet{rakshit21} proposed selecting NLS1s using $\mathrm{FWHM_{\rm Mg{\textsc{ii}}}}< 2000 \, \mathrm{km \, s^{ -1}}$.
With an Mg{\sc ii} width $\mathrm{FWHM_{Mg{\textsc{ii}}}}=1699^{+99}_{-110}$ km s$^{-1}$, J0921+0007 can be classified according to this selection criterion. \citet{rakshit21} propose a UV proxy measurement of the ratio of Fe{\sc ii} and H$\beta$ equivalent widths $r_{\rm Fe{\textsc{ii}}}$ \citep{boroson92} as $r_{\rm Fe{\textsc{ii}},UV}=\mathrm{EW(Fe{\textsc{ii}})/EW(Mg{\textsc{ii}})}$. Together with the FWHM of H$\beta$, $r_{\rm Fe\textsc{ii}}$ is one of the parameters defining the quasar main sequence \citep[e.g. see][and references therein]{marziani18}. $r_{\rm Fe{\textsc{ii}}}$ has been shown to correlate tightly with the Eddington ratio \citep[e.g.][]{rakshit17}. We computed the equivalent width of the fitted Fe{\sc ii} template in the rest-frame wavelength interval $2200-3090 \, \AA$. We note that this measurement is based on an extrapolation of the iron template, since the spectrum only covers rest frame  $\sim 2700-3150\, \AA$. We obtained $r_{\rm Fe{\textsc{ii}},UV}\sim 3.77$, a value locating the source at the strongly accreting end of the quasar main sequence (Population A see \citealt{marziani18b,marziani18}).

\section{Relative X-ray and optical/UV output}
\label{sec:x_opt}

\subsection{X-ray loudness}

\begin{figure}[]
\includegraphics[width=.5\textwidth]{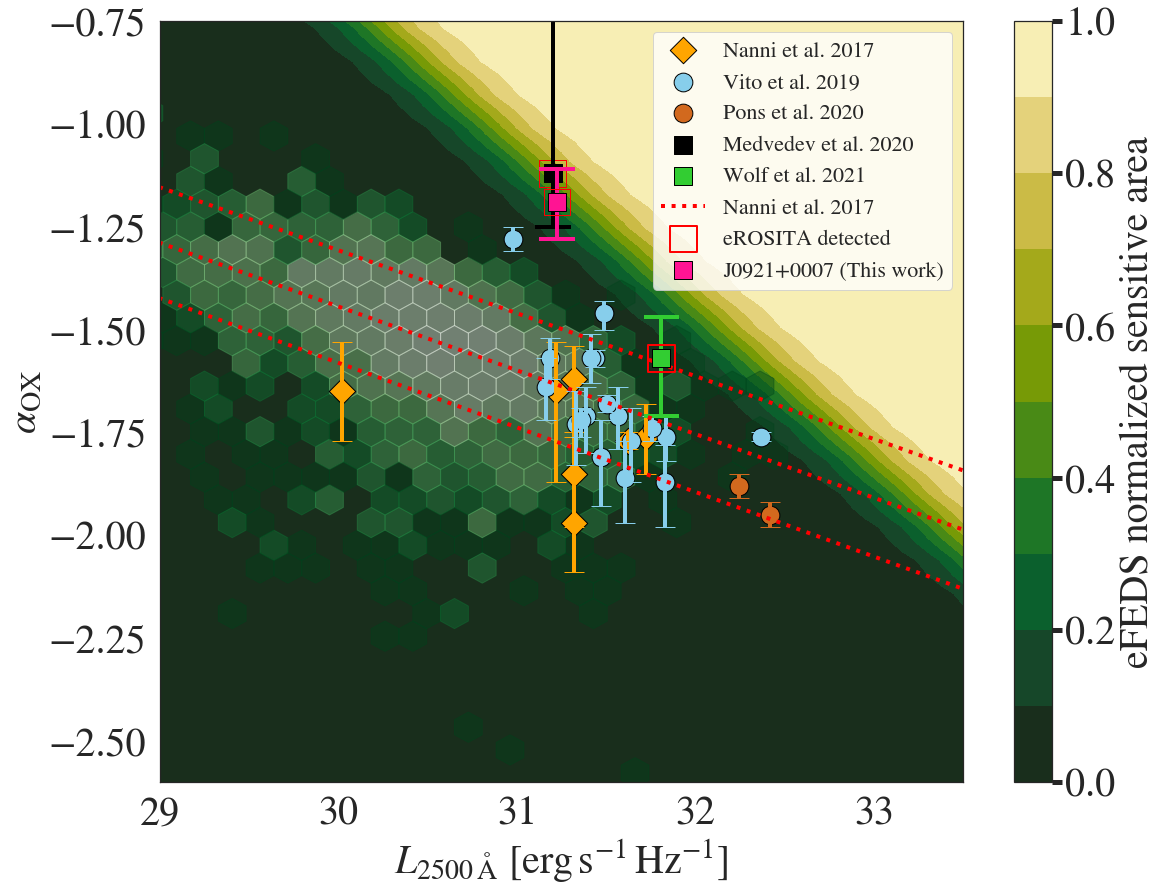}

\caption{X-ray to UV slope $\alpha_{\mathrm{OX}}$ as a function of $L_{2500}$ for quasars at $z>5.7$. Because of its strong X-ray emission, J0921+0007 (pink square) deviates significantly from the  $\alpha_{\mathrm{OX}}$-$L_{2500}$ relation, the $1\sigma$ confidence interval of which is delimited by the dotted red lines \citep{nanni17}. The sensitivity of eFEDS to sources modelled by a steep absorbed power law ($\rm \Gamma = 3$ and $N_\mathrm{H}= 3\times 10^{20} \, \mathrm{cm^{-2}}$) is traced by the colour map. The hexagonal pattern represents a sample of XMM-detected SDSS quasars presented by \citet{lusso16}.}
\label{fig:aox}
\centering
\end{figure}

The relative output of the hot corona and the UV disc emission is characterised by the ratio of monochromatic luminosities at 2 keV and 2500 $\AA$:

\begin{equation}
    \alpha_{\mathrm{OX}} = 0.384 \times \mathrm{log}\, (L_{2 \, \rm keV}/L_{2500}), 
\end{equation}
where $L_{2 \, \mathrm{keV}}$ and $L_{2500}$ are the rest-frame monochromatic luminosities at $2 \, \mathrm{keV}$ and $2500 \,  \AA$, respectively. There is a well-known anti-correlation between $\alpha_{\mathrm{OX}}$ and the $2500 \,  \AA$ monochromatic luminosity \citep[e.g.][]{steffen06,just07,lusso16}, which signifies that more UV luminous quasars tend to show a stronger UV contribution to their total emission (with respect to the coronal emission). This relation does not show evolution with redshift \citep[e.g.][]{steffen06,just07,lusso16,nanni17,vito19}. We computed the $\alpha_{\mathrm{OX}}$ by deriving $L_{2500}$ from the monochromatic luminosity of $3000 \, \AA$  and the spectral slope reported in Sect. \ref{sec:nir_sec}. The UV luminosity is $L_{2500}=(1.74 \pm 0.07) \times 10^{31} \, \mathrm{erg\,s^{-1} Hz^{-1}}$. We further obtained an estimate of $L_{2 \, \mathrm{keV}}$ from the hard broadband X-ray luminosity obtained from the spectral analysis of the \textit{Chandra} data, taking into account the posterior distribution of $\Gamma$. We obtained $\alpha_{\mathrm{OX}}=-1.21 \pm 0.09$. In Fig. \ref{fig:aox}, we show how this X-ray to UV power-law slope compares to other X-ray-detected high-redshift quasars from \citet{nanni17,vito19, pons20, medvedev21} and \citet{wolf21}. The $\alpha_{\mathrm{OX}}$ relation derived for $z>5.7$ sources by \citet{nanni17} is shown. The eFEDS normalised sensitive area to a fiducial source modelled by an absorbed power law with $\rm \Gamma = 3$ and $N_\mathrm{H}= 3\times 10^{20} \, \mathrm{cm^{-2}}$, is represented as background colour gradient. J0921+0007 shows a significantly flatter $\alpha_{\mathrm{OX}}$ slope than other high-redshift quasars with comparable UV luminosities.  Accounting for the $1\sigma$ confidence interval for the $\alpha_{\mathrm{OX}}-L_{2500}$ relation of \citet{nanni17}, the nominal $\alpha_{\mathrm{OX}}$ of J0921+0007 is a $>3\sigma$ outlier. With respect to the more conservative relation of \citet{just07}, J0921+0007 is a $>1.5\sigma$ outlier.  This indicates that the X-ray contribution to the total emission in this quasar is higher. The eFEDS sensitivity map corroborates the outlier nature in terms of relative X-ray to UV output of J0921+0007: at $L_{2500}$, we could not have detected this source in eFEDS if it followed the typical  $\alpha_{\mathrm{OX}}-L_{2500}$ relation. J0921+0007 is X-ray loud with respect to the bulk of X-ray-detected high-redshift quasars. We note that this result relies on the assumption that the steep photon index measured in the rest-frame hard band of J0921+0007, $\Gamma=3.2$, is not due to any non-coronal components (e.g. soft excess). The object of \citet{medvedev20} shows a similarly strong deviation from the $\alpha_{\mathrm{OX}}-L_{2500}$ relation, which is due to non-coronal X-ray emission from the jet. J0921+0007 in contrast is radio-quiet. The significant $\alpha_{OX}$ outlier from the \citet{vito19} sample at $\alpha_{OX}=-1.28$ and $\mathrm{log} L_{\rm 2500}\sim 31$ is the radio-quiet quasar CFHQS J1641+3755 at $z=6.04$. This source shows remarkable similarities to J0921+0007 in terms of black hole properties and its X-ray emission (see discussion).

\subsection{An increased coronal contribution to the bolometric luminosity}
\label{sec:xcig}

\begin{figure}[]
\includegraphics[width=.5\textwidth]{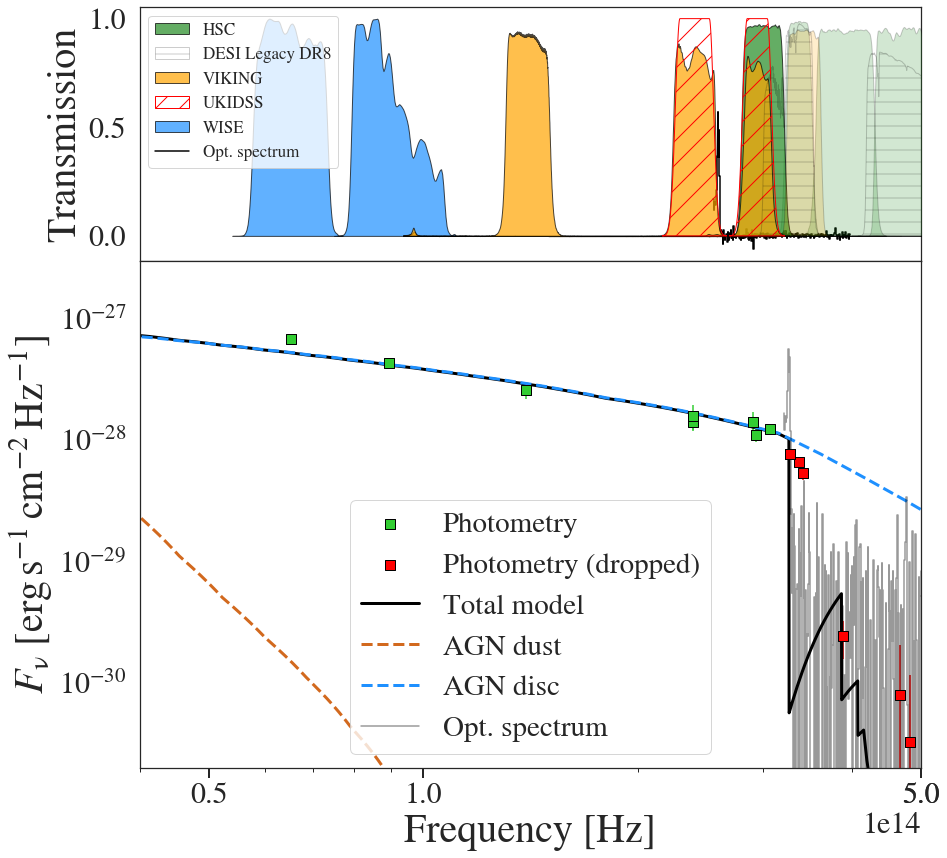}

\caption{X-CIGALE fit to the quasar optical, infrared, and X-ray photometry. Photometric points from HSC, LS8, UKIDSS, VIKING, and unWISE are marked by squares. The photometry that has been ignored for the SED fit is shown as red squares. The solid black line is the best-fitting model. The dashed blue line shows the fitted disc model and the dashed brown line the unconstrained torus emission. We also show the LDSS3 spectrum presented in Appendix \ref{sec:opt_sec}. The upper panel shows corresponding photometric filters. }
\label{fig:xcigale}
\centering
\end{figure}

In the previous section we demonstrate that the quasar shows a stronger X-ray emission than typical quasars of similar rest-frame UV luminosity $L_{2500}$. This implies that the total active galactic nucleus (AGN)  emission may be affected by the increased coronal contribution at shorter wavelengths. To account for this, we estimated a corrected AGN bolometric luminosity $L_{\rm bol,corr}$ by performing an SED fit with the tool Code Investigating GALaxy Emission (\texttt{CIGALE}; \citealt{boquien19}) and its X-ray module \texttt{X-CIGALE}, \citep{yang20}. We obtained photometry from the third public data release of the HSC-SSP, LS8, the VISTA Kilo-degree Infrared Galaxy Survey \citep[VIKING;][]{arnaboldi07}, the UKIRT Infrared Deep Sky Survey (UKIDSS; \citealt{lawrence07}), and CatWISE2020 \citep{marocco21}, by cross-matching the optical position of the quasar to these catalogues and selecting the nearest detection within $1''$. Each of the above surveys yielded a detection within $1''$.  We did not make use of photometry in the absorbed region of the spectrum at $\lambda < 1215 \AA $, since transmitted flux, for example from the quasar proximity zone, can negatively affect the fitted (see the red squares Fig. \ref{fig:xcigale}). In addition we used the 0.5-2 keV flux from the spectral fit to the \textit{Chandra} data ($\Gamma \sim 3.2$; see Chandra \texttt{BXA} in Table \ref{table:xphotometry}).  

We fitted an AGN disc model as defined in the AGN \texttt{SKIRTOR} module, which uses a disc spectral energy distribution (SED) adapted from \citet[][see also \citealt{duras17,yang20}]{feltre12}. 
We fixed the viewing angle with respect to the AGN axis to $i=30\degree$ and probed a grid of E(B-V) values for extinction in the polar direction. The AGN fraction was set to 0.999, as the host galaxy emission is completely dominated by the AGN emission in luminous, distant quasars. We further allowed a large dispersion in the $\alpha_{\mathrm{OX}}$-$L_{2500}$ relation: $\Delta \alpha_{\mathrm{OX}}= | \alpha_{\mathrm{OX}} - \alpha_{\mathrm{OX}}(L_{2500}) | = 1.0$, where $\alpha_{\mathrm{OX}}(L_{2500})$ was determined using the relation measured by \citet{just07}. We fixed the power-law photon index of the \textit{xray} module to $\Gamma=3.2$, a value supported by our X-ray spectral fitting results. We used fiducial galaxy population parameters set in the modules \textit{sfhdelayed} (delayed star-formation history) and \textit{bc03} (single stellar populations, \citealt{bruzual03}). 
Galactic dust attenuation was accounted for via
 \textit{dustatt\_calzleit} \citep{calzetti00,leitherer02}.
  The redshift was fixed to $z=6.56$. The best-fitting model has a reduced $\chi^2$ of 0.72. The AGN disc dominates the rest-frame optical/UV part of the SED. The AGN dust emission remains unconstrained. The SED as well as the total model are presented in Fig. \ref{fig:xcigale}.
 
We performed sanity checks by comparing the 2-10 X-ray luminosity $L_{2-10 \, \mathrm{keV,XCIG}}$ and the $\alpha_{\rm OX,XCIG}$ 
from the X-CIGALE output to the measurements from Sects. \ref{sec:xray_prop} and \ref{sec:nir_sec}. The values are :  $L_{2-10 \, \mathrm{keV,XCIG}}= (3.04 \pm 2.72) \times 10^{45}\, \mathrm{erg \, s^{-1}}$ and $\alpha_{\rm OX,XCIG}=-1.2$. 
These results are consistent within $1\sigma$ with the more precise measurements from the X-ray and infrared spectral analysis.

The corrected bolometric AGN luminosity was obtained by summing the intrinsic AGN disc luminosity averaged over all directions (\textit{agn.accretion\_power}) and the total X-ray luminosity (0.2-100 keV):  $L_{\rm bol,corr}=L_{\rm disc} + L_{\rm X,tot} = (9.9 \pm 4.0) \times 10^{46} \, \mathrm{erg \, s^{-1}}$. We did not include the dust emission in this calculation since it arises from re-processed nuclear UV and X-ray photons \citep[e.g.][]{lusso12,duras20}.

\begin{figure}[]
\includegraphics[width=.48\textwidth]{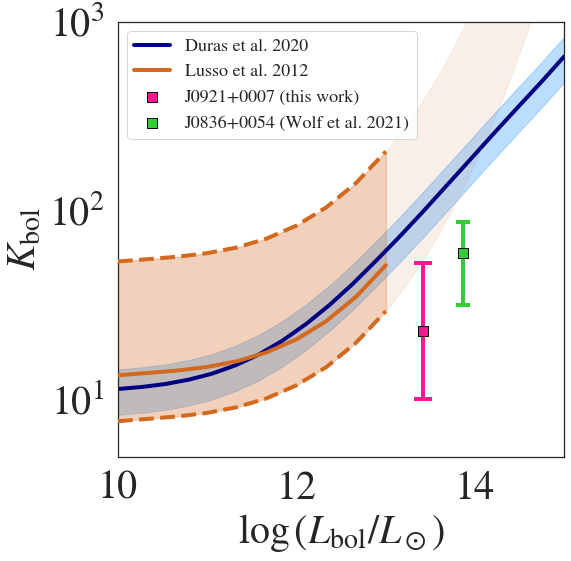}

\caption{Bolometric correction for J0921+0007 (pink error bars) compared to predictions from \citet{duras20} for the full AGN population (solid blue line) and \citet{lusso12} for type 1 AGN (solid brown line). The dispersions are quoted at the 68\% level (blue shaded area and brown shaded area). The strong X-ray emission of J0921+0007 with respect to its disc emission causes it to deviate at the $>1\sigma$ level from the typical AGN  $K_{\rm bol}-L_{\rm bol}$. The green error bars show the bolometric correction for the second eFEDS high-z quasar \citep{wolf21}. We note that the bolometric correction of \citet{lusso12} is only valid up to $\mathrm{log} \, L_{\rm bol}=13$; the extrapolation (light-shaded area) is shown for visualisation purposes only.}
\label{fig:kbol}
\centering
\end{figure}

We compared the resulting $2-10 \, \mathrm{keV}$ bolometric correction, $K_{\rm bol}=L_{\rm bol}/L_{2-10 \, \mathrm{keV}}$, to typical type 1 AGN values from literature. J0921+0007 follows a bolometric correction $K_{\rm bol}=23^{+29}_{-13}$, about a factor of 4 smaller than the prediction from the general $K_{\rm bol}-L_{\rm bol}$ relation of \citet{duras20} and the one for type 1 AGN by \citet{lusso12}, as can be seen in Fig. \ref{fig:kbol}. These findings further confirm the unusual X-ray loudness of the quasar. Using the same method, we also derived a bolometric correction for the second high-redshift quasar detected in eFEDS, SDSS J0836+0054, from the photometry presented in \citet{wolf21}. This source also shows X-ray emission excess.

\section{AGN demographics in the first gigayear of the Universe}
\label{sec:demographics}
\subsection{Comparison to XLF models}
\label{sec:xlf}
X-ray-selected quasars can be used to trace black hole accretion through cosmic time via the XLF. The AGN XLF has been studied in detail over a variety of population in deep or wide surveys  \citep{hasinger05,barger05,vito14,ueda14,buchner15,aird15,georgakakis15,fotopoulo16,ananna19}. In these works, the space density of X-ray-detected AGN has been investigated up to $z=5$ and has been shown to decline exponentially with redshift beyond the luminosity at the knee of the XLF $\mathrm{log}\, L_* \sim44$ (\citealt{brusa08,civano11}). Selecting high-redshift AGN in the \textit{Chandra} Deep Fields, \citet{vito18} extended the analysis out to $z=6$, with a sample of sources with photometric redshifts beyond $z \geq 5.5$. \citet{wolf21} showed that the eROSITA detection of a spectroscopically confirmed $z=5.81$ quasar in the eFEDS field imposes new constraints on the AGN XLF at its bright end. The detection of SDSS J0836+0054 is consistent with extrapolated models of the XLF from literature. Models with a milder decline in AGN space density 
in the highest X-ray luminosity bins are favoured by this detection. These findings are corroborated by an analysis of a sample of X-ray-detected high-redshift AGN in the ExSeSS by \citet{barlow_hall22}. We note that \citet{barlow_hall22} report the detection of ATLAS J025.6821-33.4627, a spectroscopically confirmed quasar at $z=6.31$, which had been, until now, the highest-redshift, blindly X-ray-detected AGN. J0921+0007 was blindly detected in eFEDS, albeit with a low detection-likelihood, and is therefore the highest-redshift, X-ray-selected AGN.

\begin{figure}[]
\includegraphics[width=.5\textwidth]{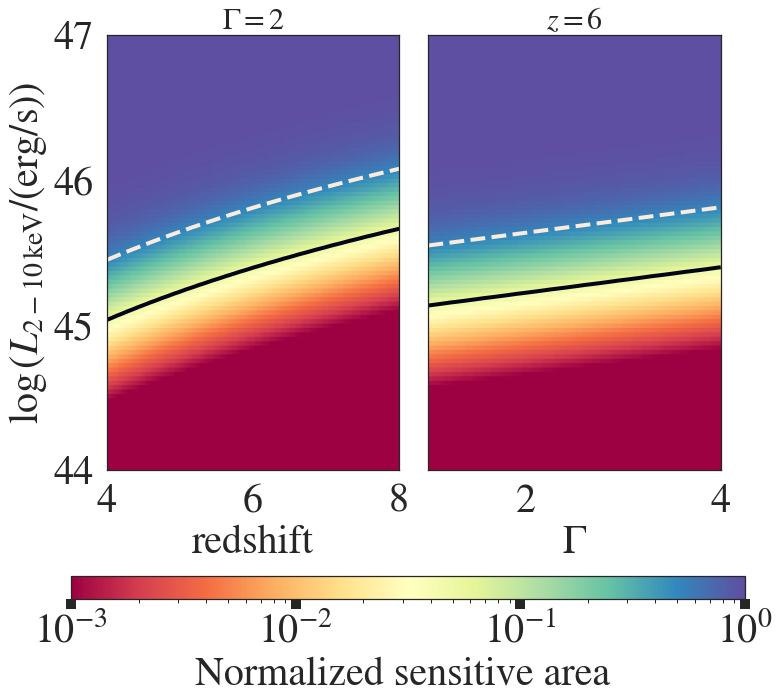}

\caption{Slices of the simulated eFEDS sensitive area cube. \textit{Left:} Sensitive area for fixed $\Gamma=2$ as a function of $ L_{2-10 \, \mathrm{keV}}$ and $z$. \textit{Right:} Sensitive area as a function of $\Gamma$ and $ L_{2-10 \, \mathrm{keV}}$ for fixed redshift $z=6$. The dotted white (solid black) line shows the parameter configurations at 50$\%$ (5$\%$) of the normalised sensitive area.}
\label{fig:sensitivits}
\centering
\end{figure}

 \citet{wolf21} show evidence against the steepest declines of the space density of luminous X-ray-detected AGN with increasing redshift. Here we go beyond this claim and show that the detection of J0921+0007 in the field is not supported by predictions of current XLF models over all luminosities extrapolated to higher redshifts.  We compare the number count predictions from the best-fitting, extrapolated XLF models presented by \cite[][luminosity-dependent density evolution]{ueda14}, \citet[][pure density evolution]{vito14}, \citet[][pure density evolution]{georgakakis15}, \citet[][luminosity-dependent density evolution]{miyaji15} and \citet[][flexible double power law]{aird15}, to the eFEDS detection. We stress that these models were evaluated on AGN samples at $z<5$ and that our comparison assumes that the parametric form of the XLF derived by these authors does not strongly evolve from $z=5$ to $z\sim 6$. We obtained number counts beyond a given redshift and luminosity threshold, $z_{\rm min}$ and $L_{\rm min}$, from the XLF models by computing

\begin{equation}
N= \int_{\mathrm{log} \, L_{\rm min}}^{\infty} \int_{z_{\rm min}}^{z=10} A_\Gamma(log \, L_\mathrm{X}, z)\frac{\mathrm{d}V}{\mathrm{d}z} \phi_m(\theta) \, \mathrm{d}\,z \, \mathrm{d}\, \mathrm{log} L_\mathrm{X}
    \label{eq:integrate}
.\end{equation}
$A_\Gamma(log \, L_\mathrm{X}, z)$ is the normalised sensitive area of the survey to a source of luminosity $L_X$ and redshift $z$, $\frac{\mathrm{d}V}{\mathrm{d}z}$ the differential comoving volume and $\phi_m(\theta)$ the model XLF (in units of $\rm Mpc^{-3}$). 
As in \citet{wolf21}, the sensitive area was obtained by converting the \texttt{apetool} \citep{georgakakis08} count-rate based area curve to a luminosity-based area curve with an X-ray spectral model. Here, we assumed a redshifted power law under Galactic absorption, (\textit{tbabs*zpowerlw}). Following this methodology, redshift-luminosity configurations as a function of the photon index were converted into soft band counts (and thereby to a normalised sensitive area) by generating X-ray spectra using the convolutional  \texttt{XSPEC} component \textit{clumin}. This conversion was fitted with logistic regression. A novelty is that we fitted the luminosity-counts relation over a grid of $\Gamma$.
 Slices of the resulting fitted surface are shown in Fig. \ref{fig:sensitivits}. In Eq. \ref{eq:integrate}, we can now evaluate the model-dependent sensitivity using this function. We integrated over all luminosities by setting $\mathrm{log} \, L_{\rm min}=42$. 

\begin{figure}[]
\centering
\includegraphics[width=0.45\textwidth]{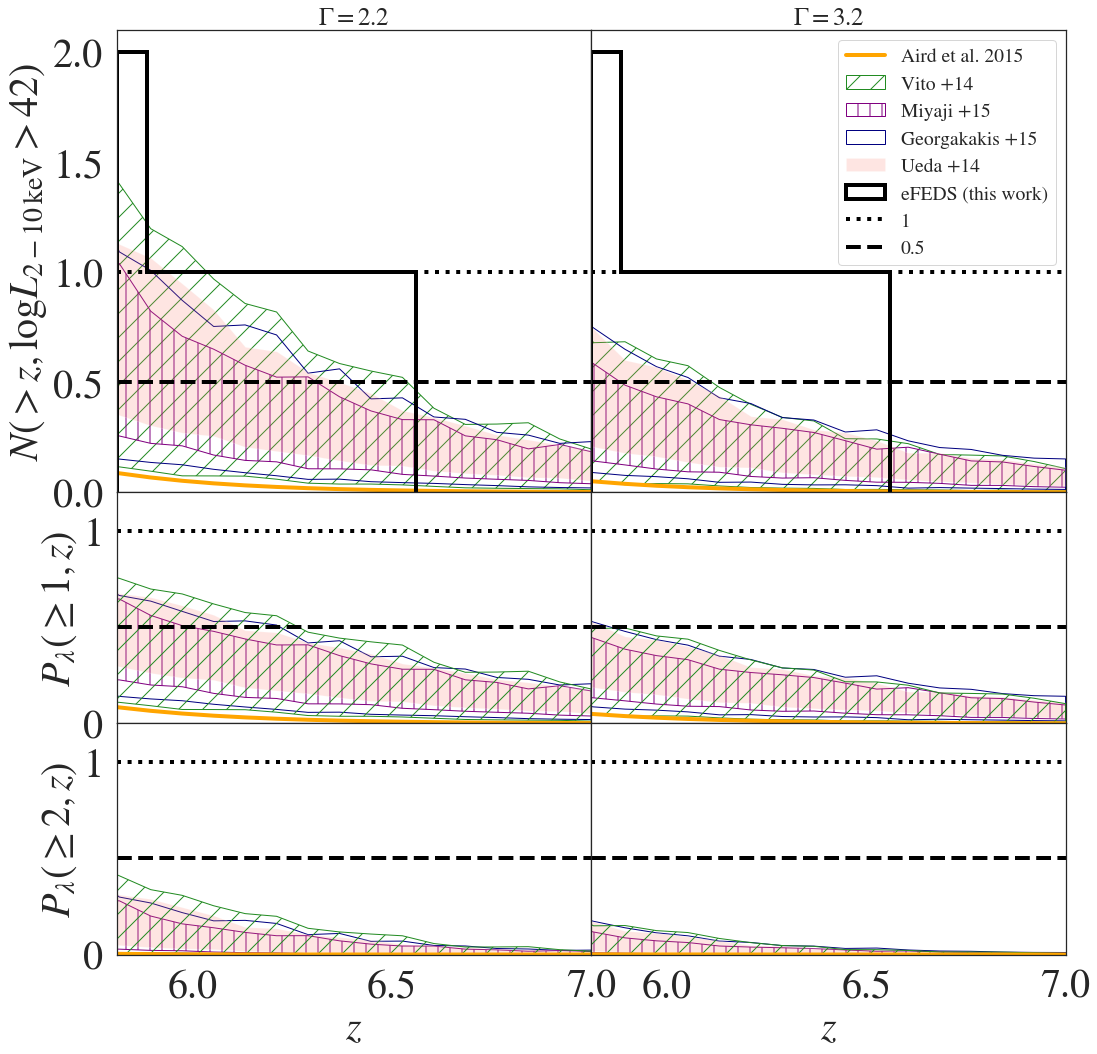}

\caption{Inverse cumulative source count predictions in eFEDS from extrapolated XLF models from \citet{vito14}, \citet{ueda14}, \citet{aird15}, \citet{miyaji15} and \citet{georgakakis15} integrated over all luminosities. The shaded and hashed areas show the $1\sigma$ confidence intervals derived from the model parameter uncertainties. These count predictions depend on the sensitive eFEDS area (Eq. \ref{eq:integrate}) and therefore on the assumed spectral model for the AGN. We show the predictions for two different photon indices, $\Gamma=2.2$ and $\Gamma=3.2$, respectively in the left and right panels. The black line shows the inverse cumulative distribution of detection in eFEDS. No model supports two detections in eFEDS for the chosen photon indices. At $\Gamma=2.2$, the \citet{vito14}, \citet{ueda14}, \citet{miyaji15} and \citet{georgakakis15} models support the unique detection at $z\geq5.81$. 
The central and lower panel present the Poisson probabilities of the XLF models supporting respectively one and two detections in eFEDS beyond a given redshift threshold. These probabilities are overall low and demonstrate the discrepancy between the eFEDS counts and the model predictions. With $\Gamma=3.2$ the detection probabilities are lower than with $\Gamma=2.2$.}
\label{fig:xlf_pred}
\centering
\end{figure}

For each of the XLF models, we evaluated Eq. \ref{eq:integrate} over the range $z_{\rm min}=5.5-7$. We accounted for two spectral models by estimating the sensitive survey area  $A(\mathrm{log} \, L_\mathrm{X}, z)$ for $\Gamma=2.2$ \citep{vito19} and $\Gamma=3.2$, the median value derived from the \textit{Chandra} observation of J0921+0007. This effectively yields the expected number counts over all luminosities beyond an increasing redshift threshold $z_{\rm min}$. For each model and each $z_{\rm min}$ we computed the 15.9-th and 84.1-th percentile of the count expectations to estimate their $1\sigma$ confidence intervals. The resulting inverse cumulative distribution of predicted counts is presented in Fig. \ref{fig:xlf_pred}. We obtained $1\sigma$ confidence intervals by sampling from the parameter uncertainties of the models. As the covariance matrix of these parameters was not accounted for, the uncertainty may have been over-estimated.
The model predictions are compared to the source counts detected in eFEDS (Fig. \ref{fig:xlf_pred}). We  can conclude that over all luminosities, none of the extrapolated XLF models supports two detections 
in the eFEDS field at high redshift, regardless of the X-ray spectral model. At $z>5.81$, the models from \cite{vito14}, \citet{ueda14}, \citet{miyaji15} and \citet{georgakakis15} 
are consistent with one detection to within $1\sigma$. No model supports a detection at $z=6.56$. This is further shown in the 2 lower panels of Fig. \ref{fig:xlf_pred}, where we present for each lower-redshift interval edge, $z_{\rm min}$, the Poisson probability for each XLF model of supporting at least one ($P_\lambda (\geq 1,z)$) or two counts ($P_\lambda (\geq 2,z)$)in the eFEDS field. Even for the model with the highest number count prediction at $z\geq 5.81$ \citep{vito14} the Poisson probability of detecting two sources in eFEDS is less than 0.4. For all models, the probability of generating 1 count at $z \geq 6.56$ is less than 0.5. As expected for sources with steeper photon indices, the count expectations are lower. In particular, a detection at $z \geq 6.56$ is unlikely with $P_\lambda (\geq 1,z\geq 6.56) < 0.25$. As discussed in \citet{wolf21}, the volume probed by the eFEDS survey at $z=5.81-6.56$ is such that the uncertainty due to cosmic variance in the expected source counts is negligible with respect to the Poisson error. We conclude that the current extrapolated XLF models under-predict the number of high-redshift quasars that we detect in eFEDS. 

\subsection{Contribution of X-ray-luminous quasars to accretion density at $z\sim 6$}

\begin{figure*}[]
\centering
\includegraphics[width=.85\textwidth]{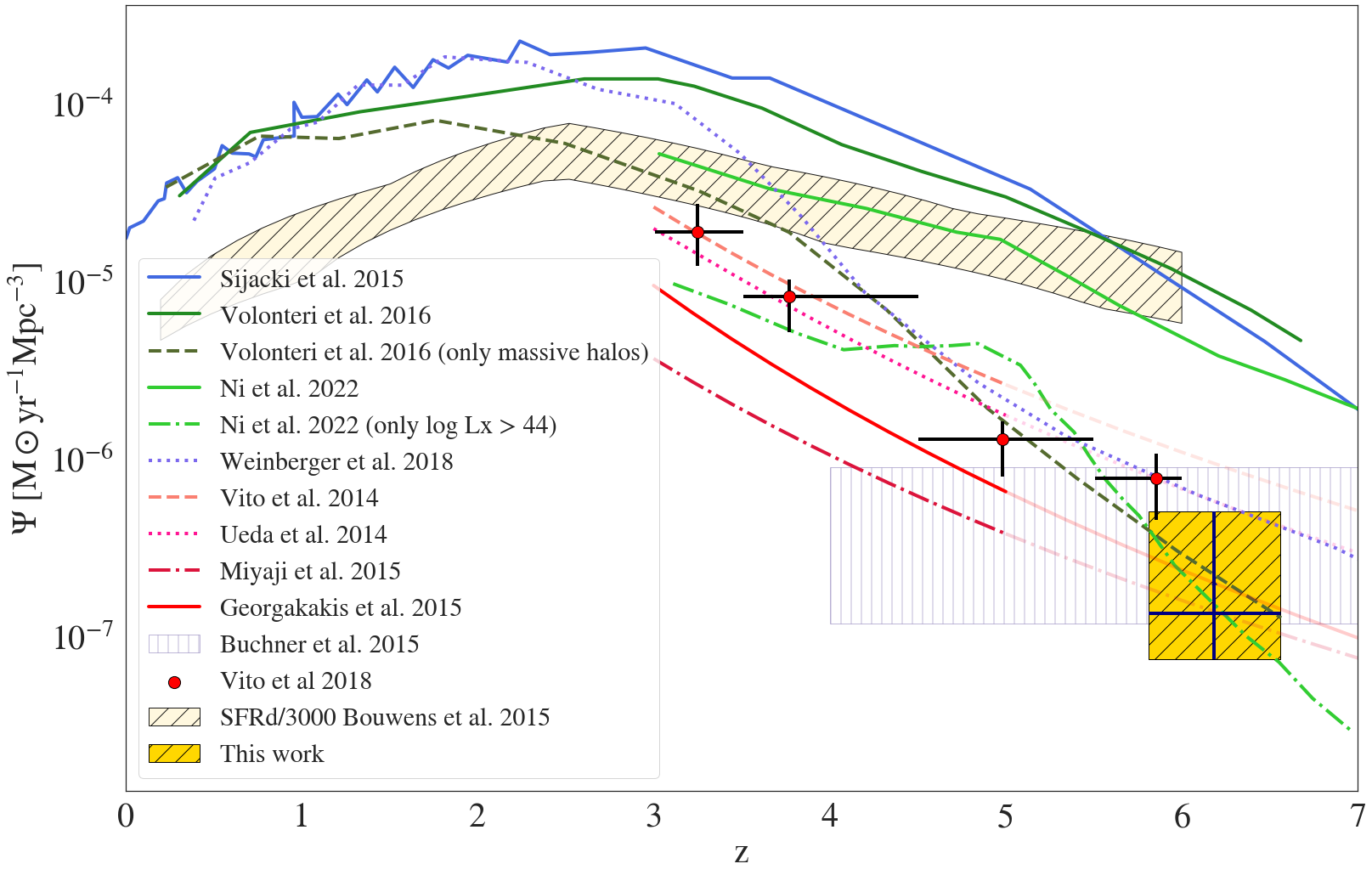}

\caption{Black hole accretion rate density (BHAD) for various XLF models and growth simulations. XLF models from \cite{vito14}, \cite{ueda14}, \cite{miyaji15} and \cite{georgakakis15} are extrapolated beyond $z=5$ (pale continuation of the red BHAD curves), and the prediction from \citet{buchner15} is given over the full $z=4-7$ range.  We include observational results from \citet{vito18} from the \textit{Chandra} Deep Fields. The measurement derived from the high-redshift quasar detections in eFEDS is shown as a yellow square. Our result is consistent with theoretical predictions restricted to the highest halo (and black hole) masses \citep{volonteri16,yueying22}. For comparison, a scaled version of the star formation rate density from \citet{bouwen15} is shown as beige-shaded and dashed area.}
\label{fig:bhad}
\centering
\end{figure*}

The total black hole mass accretion rate per unit volume can be traced through cosmic time via the black hole accretion rate density (BHAD), $\Psi_{\rm bhar}(z)$, which is related to the AGN bolometric luminosity function $\phi (L_{\rm bol},z)$, as
\begin{equation}
 \Psi_{\rm bhar}(z) = \int \frac{1-\epsilon}{\epsilon c^2} L_{\rm bol} \, \phi (L_{\rm bol},z) \mathrm{d}\, \mathrm{log} L_{\rm bol},
\end{equation}
where $L_{\rm bol}$ is the bolometric luminosity and $\epsilon$ the radiative efficiency. The radiative efficiency $\epsilon$ is related to accretion efficiency $\eta$ and the physics governing the accretion flow. We assume that the X-ray-luminous eFEDS quasars accrete above the critical rate, which delimits a radiatively efficient accretion disc from an inefficient one. We adopted a standard thin disc estimate $\epsilon= \eta=0.1$ \citep[e.g.][]{soltan82,fabian99,merloni08,delvechhio14}, that is, the radiative efficiency is set equal to the accretion efficiency. We note that the value of $\epsilon$ only affects the normalisation of $\Psi_{\rm bhar}$. 
At higher redshifts, the evolution of the BHAD has been derived from X-ray-detected \citep{aird15,vito18} and X-ray-undetected, stacked AGN \citep{vito16}. To estimate the total contribution of the eFEDS-detected quasars, we first note that
\begin{equation}
\label{eq:bhad_xfl}
 \Psi_{\rm bhar}(z) = \frac{1-\epsilon K_{\rm bol}}{\epsilon c^2} \int L_{\rm 2-10 \, keV} \, \phi_X (L_{\rm 2-10 \, keV},z) \mathrm{d}\, \mathrm{log} L_{\rm 2-10 \, keV},
\end{equation}
where $L_{\rm 2-10 \, keV}$ is the 2-10 keV luminosity, $\phi_X (L_{\rm 2-10 \, keV},z)$ the hard XLF and $K_{\rm bol}$ the bolometric correction from the 2-10 keV band. The integral is the total AGN emissivity per unit volume. Therefore, we can rewrite Eq. 5 as 
\begin{equation}
 \Psi_{\rm bhar}(z)  \approx  \frac{1-\epsilon}{\epsilon c^2} \times \sum_{\Delta z} \frac{L_{\rm bol,AGN}}{V_{\rm eFEDS, \Gamma}},
 \label{eq:bhad_emissivity}
\end{equation}
where $L_{\rm bol,AGN}$ the bolometric AGN luminosity of sources detected in the $\Delta z$ bin and $V_{\rm eFEDS, \Gamma}$ the sensitive comoving volume of the eFEDS survey. The sum is taken over all detections in $\Delta z$. We computed this estimator for the redshift interval $z=[5.81-6.56]$, the redshift interval spanned by the two quasars. 
The contribution of the eFEDS high-z quasars to black hole accretion in this redshift bin can be obtained by summing up the ratio of accretion luminosities obtained in the SED fits presented in Sect. \ref{sec:xcig} (\textit{agn.accretion\_power}) to the corresponding eFEDS sensitive volume \citep{vito16}. We obtained this volume by accounting for the sensitivity to sources that have 2-10 keV luminosities of the quasars detected in eFEDS (see Fig. \ref{fig:sensitivits}). We assumed $\Gamma=2.2$ for the sensitive survey area of J0836+0054 \citep{wolf21} and $\Gamma=3.2$ for J0921+0007 (as derived from the spectral fit in Sec \ref{sec:xray_prop}). The resulting accretion density is $\Psi_{z=5.81-6.56}=(1.36^{+3.58}_{-0.62}) \times 10^{-7} \mathrm{M_\odot\, yr^{-1}Mpc^{-3}}$. We note that this total AGN emissivity per unit comoving volume only accounts for the un-extincted disc luminosity and not the disc photons reprocessed by the torus and the corona.

We compare this result to lower-redshift measurements and theoretical predictions in Fig. \ref{fig:bhad}.  The theoretical predictions shown in this figure assume different seeding masses and growth modes. We show results across the entire halo mass scale \citep{sijacki15,volonteri16,Weinberger18,yueying22} and results restricted to sub-samples at the high-mass and high-luminosity end \citep{volonteri16,yueying22}. The departure from co-evolution of black hole accretion rate and the star formation rate at $z>3$ is in general difficult to achieve with a cosmological model.

It has been suggested that good agreement between observations and simulations is only warranted when only including large black hole or halo masses, while using all masses causes simulations to over-predict the black hole accretion density \citep{sijacki15,volonteri16}. For example, the results from the Horizon-AGN simulations \citep{volonteri16} are presented for the total mass range (solid dark green) and for halos with a halo hole mass of $>5 \times 10^{11} \mathrm{M_\odot}$ (dashed dark green). \citet{volonteri16} proposed that supernova feedback could be the reason why observational results only agree with simulations when applying a high-mass threshold. Indeed supernovae feedback is expected to deplete the AGN core, effectively stopping black hole growth in low-mass galaxies \citep{dubois15, habouzit17}. As the galaxy grows in mass its deeper gravitational potential allows it to more efficiently confine the gas in the nucleus. The black hole accretion density derived from the X-ray-luminous eFEDS detections is in perfect agreement with the predictions from \citet{volonteri16} at $M_{\rm halo}>5 \times 10^{11} \mathrm{M_\odot}$. Similarly, in their ASTRID simulations
\citet{yueying22} present their BHAD for various X-ray luminosity thresholds. At high luminosities (log $L_X > 44$) the predictions of the steeper falling BHAD curves and the value derived from eFEDS are in excellent agreement. However, we stress again that the eFEDS survey is not sensitive to $44<\mathrm{log }\,L_X<45$ in the probed redshift regime. We are therefore missing contributions to accretion rate density from quasars accounted for in the \citet{yueying22} BHAD curve.
Because of its sensitivity limit, eFEDS becomes highly incomplete at log $L_X< 45$ at $z\sim 6$ (see Fig. \ref{fig:sensitivits}). The sample of \citet{vito18} is extracted from the \textit{Chandra} Deep Fields. These surveys are smaller in area but deeper than eFEDS and can sample AGN efficiently down to luminosities log $L_X > 42.5$. This difference in sensitivity explains the discrepancy seen at $z \sim 6$ between the results of \citet{vito18} and the lower boundary obtained in this work: eFEDS misses quasars in the range $42.5<L_X<45$, which still significantly contribute to black hole accretion. We also point out that the main and supplementary eFEDS catalogues are not spectroscopically complete. In this regard, the data point we derived should be considered a lower limit on the BHAD. 
Another source of discrepancy between the BHAD derived from various X-ray surveys is the use of photometric redshifts, which can potentially populate the $z>5$ bin with interlopers. For our study, we only used X-ray sources with clear multi-wavelength identifications and spectroscopic redshifts. Following \citet{volonteri16}, our results suggest that, assuming supernova-feedback-regulated black hole growth, most black hole accretion is dominated by extremely luminous AGN. Alternatively, the agreement with the prediction of \citet{yueying22} indicates that at log $L_x > 44$, black hole accretion is truly dominated by the most X-ray-luminous quasars at log $L_X>45$.

We have shown that the extrapolated XLF models by \citet{vito14}, \citet{ueda14}, \citet{aird15}, \citet{miyaji15} and \citet{georgakakis15} underestimate the number of high-z quasar detections in eFEDS (see Fig. \ref{fig:xlf_pred}); however, it can be seen in Fig. \ref{fig:bhad} that the black hole accretion density derived from these models appears to be consistent with the one resulting from the eFEDS detections. This can be explained by the high X-ray to optical flux ratio for both eFEDS quasars, which results in significantly smaller bolometric corrections (see Fig. \ref{fig:kbol}). The bolometric correction assumed for the conversion of XLFs to BHAD (Eq. \ref{eq:bhad_xfl}) from \cite{duras20} causes a higher extrapolated BHAD, despite the underprediction of actual luminous high-z sources in the field. In addition, the black hole accretion density is calculated in Eq. \ref{eq:bhad_emissivity} as the efficiency-scaled total emissivity of the quasars detected in the $z=5.81-6.56$ interval and is therefore inversely proportional to the sensitive volume probed by eFEDS at the luminosities and redshifts of these quasars. The X-ray luminosity-redshift configurations of the quasars detected in eFEDS, in particular that of SDSS J0836+0054, result in a larger sensitive volume (see Fig. \ref{fig:lx_plot}) and therefore a lower contribution to the black hole accretion density.


\section{Discussion and conclusions}

We have characterised a $z>6$ super-Eddington-accreting NLS1 with low black hole mass based on archival photometry and a new NIR spectrum. We discuss how our findings support the idea that $z > 6$ NLS1s potentially show physical properties that resemble those of their lower-redshift counterparts. At $z=6.56$, J0921+0007 is the most distant X-ray-selected AGN to date and can therefore be used to impose constraints on the high-z XLF.

We derived a comparatively low black hole mass \citep[for a sample of high-redshift quasars with comparable optical/UV luminosity, see e.g.][]{onoue19}, which implies that the source is accreting at a super-Eddington rate. The values reported in this work ($M_\mathrm{BH} = (2.5 \pm 0.3) \times 10^8 M_\odot$ and $\lambda = 2.3^{+0.4}_{-0.3}$) are consistent with the typical properties of local NLS1s \citep[e.g.][]{sulentic00,collin04,rakshit17}. 
We obtained a relatively steep power-law fit to the X-ray spectrum of the source: $\Gamma = 3.2$. Such a high value is usually found in the rest-frame soft band of archetypal low-z NLS1s \citep[e.g.][]{boller96,brandt97,ohja20}. In the rest-frame hard band, NLS1s typically show photon indices below this value \citep[$\sim 2$; e.g.][]{zhou10}. The steeper photon index found here can  be driven by either the large accretion rate \citep{shemmer06} or the presence of unresolved non-coronal components. Similar sources, in terms of rest-frame optical properties, have been discovered by \citet{koptelova17} and \citet{banados21}. The quasar CFHQS J1641+3755 at $z=6.04$ was initially discovered by \citet{willott07}. \citet{willott10} obtained NIR spectroscopy for this source with the NIRI instrument on the Gemini-North Telescope. It shows an Mg{\sc ii} profile ($\mathrm{FWHM_{\rm Mg{\textsc{ii}}}} = 1740 \pm 190 \, \mathrm{km\, s^{-1}}$) that is very similar to the one observed in the MOIRCS spectrum of J0921+0007 presented in our work. According to the \citet{rakshit21} classification criterion, this makes it a high-z NLS1.  The derived black hole mass and Eddington ratio are $M_{\rm bh}=2.4 \times 10^{8} \, M_\odot$ and $\lambda= 2.3$, indicating that CFHQS J1641+3755 may be powered by a low-mass, strongly accreting black hole. \citet{vito19} report the X-ray observation of this quasar with \textit{Chandra}. While it has a relatively modest bolometric luminosity, it is the second-most X-ray-luminous source in their sample, making it deviate from the $\alpha_{OX}-L_{\rm UV}$ by 1.8$\sigma$ with respect to the best-fitting relation of \citet{steffen06}. \citet{vito19} also derive a steep photon index for this source ($\Gamma=2.56$). We conclude that CFHQS J1641+3755 is another archetypal NLS1 at high redshift. To further support the NLS1 classification of J0921+0007, we measured the extent of the quasar proximity zone and present these results in Appendix \ref{sec:opt_sec}.

 In Sect. \ref{sec:xlf} we show that the number of high-z source detections in the eFEDS field, combining the present work with the results from \citet{wolf21}, is significantly higher than predictions from a large range of XLF models in the literature extrapolated out to $z\sim 6$. eFEDS is the largest contiguous public X-ray survey to date with sufficient depth to investigate $z\sim 6$ AGN demographics. It probes a cosmological volume that is  sufficiently large to contain rare rare $\mathrm{log} \, L_\mathrm{X}> 45$ quasars at high redshift, including the unexpected class of high-z NLS1s discussed in this work. The discrepancy between previous XLF models obtained using smaller pencil-beam or non-contiguous surveys underlines the necessity for wide surveys to obtain a realistic census of the rare, powerful sources at the bright end of the XLF \citep[see e.g.][]{barlow_hall22}.

Stacking the \textit{Chandra} Deep Field South data from a sample of $3.5<z<6.5$ galaxies, \citet{vito16} show that the contribution of detected luminous quasars at $z\sim 6$ to the black hole accretion density is higher than the one from stacked undetected sources by an order of magnitude. These findings corroborate the results of \citet{volonteri16}, who concluded that most of the black hole growth is supplemented by luminous quasars ($L_{\rm bol}>10^{43} \, \mathrm{erg\, s^{-1}}$) in massive halos ($>5\times 10^{11} \mathrm{M_\odot}$). The accretion density derived from the two detected quasars in eFEDS is consistent with these previous results. At the flux limit of eFEDS, it is only possible to sample the ultra-luminous population, $L_{\rm 2-10 \, keV}>10^{45} \, \mathrm{erg\, s^{-1}}$. Despite this sensitivity limit, our results are already consistent with the predictions from \citet{volonteri16} and \citet{yueying22}, indicating that most of the black hole growth is in fact driven by X-ray-ultra-luminous quasars, above the eROSITA sensitivity limit. J0921+0007 is an unexpected member of this category of extreme quasars. Its X-ray luminosity is significantly higher than the value extrapolated from the $\alpha_{\rm OX}-L_{\rm UV}$ relation. In order to quantify how much of the accretion density is in fact driven by young, super-Eddington black holes, a wider survey area will be required at this depth to obtain a more informative sample. This will be made possible in the cumulative eROSITA All-Sky Survey \citep[][see also \citealt{seppi22}]{merloni12}.

\begin{acknowledgements}

    JW acknowledges support by the Deutsche Forschungsgemeinschaft (DFG, German Research Foundation) under Germany's Excellence Strategy - EXC-2094 - 390783311. He would also like to thank Peter Predehl for his help with the observation presented in this work.
    
    \newline
    
     MO acknowledges support by the Natural Science Foundation of China (12150410307). MB is supported by the European  Innovative Training  Network  (ITN)  "BiD4BEST" funded by  the  Marie  Sklodowska-Curie Actions in Horizon 2020 (GA 860744).
     
     \newline
    
    This work is based on data from eROSITA, the primary instrument aboard SRG, a joint Russian-German science mission supported by the Russian Space Agency (Roskosmos), in the interests of the Russian Academy of Sciences represented by its Space Research Institute (IKI), and the Deutsches Zentrum für Luft- und Raumfahrt (DLR). The SRG spacecraft was built by Lavochkin Association (NPOL) and its subcontractors, and is operated by NPOL with support from the Max Planck Institute for Extraterrestrial Physics (MPE).
    
    The development and construction of the eROSITA X-ray instrument was led by MPE, with contributions from the Dr. Karl Remeis Observatory Bamberg \& ECAP (FAU Erlangen-Nuernberg), the University of Hamburg Observatory, the Leibniz Institute for Astrophysics Potsdam (AIP), and the Institute for Astronomy and Astrophysics of the University of Tübingen, with the support of DLR and the Max Planck Society. The Argelander Institute for Astronomy of the University of Bonn and the Ludwig Maximilians Universität Munich also participated in the science preparation for eROSITA.
    
    The eROSITA data shown here were processed using the eSASS software system developed by the German eROSITA consortium.
     \newline

  The scientific results reported in this article are based to a significant degree on observations made by the Chandra X-ray Observatory. 
       \newline
This research is based on data collected at the Subaru Telescope, which is operated by the National Astronomical Observatory of Japan. We are honored and grateful for the opportunity of observing the Universe from Maunakea, which has the cultural, historical, and natural significance in Hawaii. 
 \newline

    The Legacy Surveys consist of three individual and complementary projects: the Dark Energy Camera Legacy Survey (DECaLS; NOAO Proposal ID $\#$ 2014B-0404; PIs: David Schlegel and Arjun Dey), the Beijing-Arizona Sky Survey (BASS; NOAO Proposal ID $\#$ 2015A-0801; PIs: Zhou Xu and Xiaohui Fan), and the Mayall z-band Legacy Survey (MzLS; NOAO Proposal ID $\#$ 2016A-0453; PI: Arjun Dey). DECaLS, BASS and MzLS together include data obtained, respectively, at the Blanco telescope, Cerro Tololo Inter-American Observatory, National Optical Astronomy Observatory (NOAO); the Bok telescope, Steward Observatory, University of Arizona; and the Mayall telescope, Kitt Peak National Observatory, NOAO. The Legacy Surveys project is honored to be permitted to conduct astronomical research on Iolkam Du'ag (Kitt Peak), a mountain with particular significance to the Tohono O'odham Nation.

NOAO is operated by the Association of Universities for Research in Astronomy (AURA) under a cooperative agreement with the National Science Foundation.

This project used data obtained with the Dark Energy Camera (DECam), which was constructed by the Dark Energy Survey (DES) collaboration. Funding for the DES Projects has been provided by the U.S. Department of Energy, the U.S. National Science Foundation, the Ministry of Science and Education of Spain, the Science and Technology Facilities Council of the United Kingdom, the Higher Education Funding Council for England, the National Center for Supercomputing Applications at the University of Illinois at Urbana-Champaign, the Kavli Institute of Cosmological Physics at the University of Chicago, Center for Cosmology and Astro-Particle Physics at the Ohio State University, the Mitchell Institute for Fundamental Physics and Astronomy at Texas A\&M University, Financiadora de Estudos e Projetos, Fundacao Carlos Chagas Filho de Amparo, Financiadora de Estudos e Projetos, Fundacao Carlos Chagas Filho de Amparo a Pesquisa do Estado do Rio de Janeiro, Conselho Nacional de Desenvolvimento Cientifico e Tecnologico and the Ministerio da Ciencia, Tecnologia e Inovacao, the Deutsche Forschungsgemeinschaft and the Collaborating Institutions in the Dark Energy Survey. The Collaborating Institutions are Argonne National Laboratory, the University of California at Santa Cruz, the University of Cambridge, Centro de Investigaciones Energeticas, Medioambientales y Tecnologicas-Madrid, the University of Chicago, University College London, the DES-Brazil Consortium, the University of Edinburgh, the Eidgenossische Technische Hochschule (ETH) Zurich, Fermi National Accelerator Laboratory, the University of Illinois at Urbana-Champaign, the Institut de Ciencies de l'Espai (IEEC/CSIC), the Institut de Fisica d'Altes Energies, Lawrence Berkeley National Laboratory, the Ludwig-Maximilians Universitat Munchen and the associated Excellence Cluster Universe, the University of Michigan, the National Optical Astronomy Observatory, the University of Nottingham, the Ohio State University, the University of Pennsylvania, the University of Portsmouth, SLAC National Accelerator Laboratory, Stanford University, the University of Sussex, and Texas A\&M University.

BASS is a key project of the Telescope Access Program (TAP), which has been funded by the National Astronomical Observatories of China, the Chinese Academy of Sciences (the Strategic Priority Research Program "The Emergence of Cosmological Structures" Grant \# XDB09000000), and the Special Fund for Astronomy from the Ministry of Finance. The BASS is also supported by the External Cooperation Program of Chinese Academy of Sciences (Grant \# 114A11KYSB20160057), and Chinese National Natural Science Foundation (Grant \# 11433005).

The Legacy Survey team makes use of data products from the Near-Earth Object Wide-field Infrared Survey Explorer (NEOWISE), which is a project of the Jet Propulsion Laboratory/California Institute of Technology. NEOWISE is funded by the National Aeronautics and Space Administration.

The Legacy Surveys imaging of the DESI footprint is supported by the Director, Office of Science, Office of High Energy Physics of the U.S. Department of Energy under Contract No. DE-AC02-05CH1123, by the National Energy Research Scientific Computing Center, a DOE Office of Science User Facility under the same contract; and by the U.S. National Science Foundation, Division of Astronomical Sciences under Contract No. AST-0950945 to NOAO.

\newline

The Hyper Suprime-Cam (HSC) collaboration includes the astronomical communities of Japan and Taiwan, and Princeton University. The HSC instrumentation and software were developed by the National Astronomical Observatory of Japan (NAOJ), the Kavli Institute for the Physics and Mathematics of the Universe (Kavli IPMU), the University of Tokyo, the High Energy Accelerator Research Organization (KEK), the Academia Sinica Institute for Astronomy and Astrophysics in Taiwan (ASIAA), and Princeton University. Funding was contributed by the FIRST program from Japanese Cabinet Office, the Ministry of Education, Culture, Sports, Science and Technology (MEXT), the Japan Society for the Promotion of Science (JSPS), Japan Science and Technology Agency (JST), the Toray Science Foundation, NAOJ, Kavli IPMU, KEK, ASIAA, and Princeton University.  
 
\newline

    This research made use of Astropy,\footnote{http://www.astropy.org} a community-developed core Python package for Astronomy \citep{astropy13, astropy18}. 
    \newline
      This research has made use of software provided by the Chandra X-ray Center (CXC) in the application packages CIAO and Sherpa.
    \newline
    In addition this research made use of BXA (https://johannesbuchner.github.io/BXA/) and the \texttt{corner} package \citep{corner}.

\end{acknowledgements}

\bibliographystyle{aa}
\bibliography{bibliography}

\begin{thebibliography}{153}
\expandafter\ifx\csname natexlab\endcsname\relax\def\natexlab#1{#1}\fi

\bibitem[{{Aihara} {et~al.}(2022){Aihara}, {AlSayyad}, {Ando}, {Armstrong},
  {Bosch}, {Egami}, {Furusawa}, {Furusawa}, {Harasawa}, {Harikane}, {Hsieh},
  {Ikeda}, {Ito}, {Iwata}, {Kodama}, {Koike}, {Kokubo}, {Komiyama}, {Li},
  {Liang}, {Lin}, {Lupton}, {Lust}, {MacArthur}, {Mawatari}, {Mineo},
  {Miyatake}, {Miyazaki}, {More}, {Morishima}, {Murayama}, {Nakajima},
  {Nakata}, {Nishizawa}, {Oguri}, {Okabe}, {Okura}, {Ono}, {Osato}, {Ouchi},
  {Pan}, {Plazas Malag{\'o}n}, {Price}, {Reed}, {Rykoff}, {Shibuya},
  {Simunovic}, {Strauss}, {Sugimori}, {Suto}, {Suzuki}, {Takada}, {Takagi},
  {Takata}, {Takita}, {Tanaka}, {Tang}, {Taranu}, {Terai}, {Toba}, {Turner},
  {Uchiyama}, {Vijarnwannaluk}, {Waters}, {Yamada}, {Yamamoto}, \&
  {Yamashita}}]{aihara22}
{Aihara}, H., {AlSayyad}, Y., {Ando}, M., {et~al.} 2022, \pasj, 74, 247

\bibitem[{{Aird} {et~al.}(2015){Aird}, {Coil}, {Georgakakis}, {Nandra},
  {Barro}, \& {P{\'e}rez-Gonz{\'a}lez}}]{aird15}
{Aird}, J., {Coil}, A.~L., {Georgakakis}, A., {et~al.} 2015, \mnras, 451, 1892

\bibitem[{{Ananna} {et~al.}(2019){Ananna}, {Treister}, {Urry}, {Ricci},
  {Kirkpatrick}, {LaMassa}, {Buchner}, {Civano}, {Tremmel}, \&
  {Marchesi}}]{ananna19}
{Ananna}, T.~T., {Treister}, E., {Urry}, C.~M., {et~al.} 2019, \apj, 871, 240

\bibitem[{{Arnaboldi} {et~al.}(2007){Arnaboldi}, {Neeser}, {Parker}, {Rosati},
  {Lombardi}, {Dietrich}, \& {Hummel}}]{arnaboldi07}
{Arnaboldi}, M., {Neeser}, M.~J., {Parker}, L.~C., {et~al.} 2007, The
  Messenger, 127, 28

\bibitem[{{Arnaud}(1996)}]{arnaud96}
{Arnaud}, K.~A. 1996, in Astronomical Society of the Pacific Conference Series,
  Vol. 101, Astronomical Data Analysis Software and Systems V, ed. G.~H.
  {Jacoby} \& J.~{Barnes}, 17

\bibitem[{{Arnaud} {et~al.}(1985){Arnaud}, {Branduardi-Raymont}, {Culhane},
  {Fabian}, {Hazard}, {McGlynn}, {Shafer}, {Tennant}, \& {Ward}}]{arnaud85}
{Arnaud}, K.~A., {Branduardi-Raymont}, G., {Culhane}, J.~L., {et~al.} 1985,
  \mnras, 217, 105

\bibitem[{{Astropy Collaboration} {et~al.}(2018){Astropy Collaboration},
  {Price-Whelan}, {Sip{\H{o}}cz}, {G{\"u}nther}, {Lim}, {Crawford}, {Conseil},
  {Shupe}, {Craig}, {Dencheva}, {Ginsburg}, {VanderPlas}, {Bradley},
  {P{\'e}rez-Su{\'a}rez}, {de Val-Borro}, {Aldcroft}, {Cruz}, {Robitaille},
  {Tollerud}, {Ardelean}, {Babej}, {Bach}, {Bachetti}, {Bakanov}, {Bamford},
  {Barentsen}, {Barmby}, {Baumbach}, {Berry}, {Biscani}, {Boquien}, {Bostroem},
  {Bouma}, {Brammer}, {Bray}, {Breytenbach}, {Buddelmeijer}, {Burke},
  {Calderone}, {Cano Rodr{\'\i}guez}, {Cara}, {Cardoso}, {Cheedella}, {Copin},
  {Corrales}, {Crichton}, {D'Avella}, {Deil}, {Depagne}, {Dietrich}, {Donath},
  {Droettboom}, {Earl}, {Erben}, {Fabbro}, {Ferreira}, {Finethy}, {Fox},
  {Garrison}, {Gibbons}, {Goldstein}, {Gommers}, {Greco}, {Greenfield},
  {Groener}, {Grollier}, {Hagen}, {Hirst}, {Homeier}, {Horton}, {Hosseinzadeh},
  {Hu}, {Hunkeler}, {Ivezi{\'c}}, {Jain}, {Jenness}, {Kanarek}, {Kendrew},
  {Kern}, {Kerzendorf}, {Khvalko}, {King}, {Kirkby}, {Kulkarni}, {Kumar},
  {Lee}, {Lenz}, {Littlefair}, {Ma}, {Macleod}, {Mastropietro}, {McCully},
  {Montagnac}, {Morris}, {Mueller}, {Mumford}, {Muna}, {Murphy}, {Nelson},
  {Nguyen}, {Ninan}, {N{\"o}the}, {Ogaz}, {Oh}, {Parejko}, {Parley}, {Pascual},
  {Patil}, {Patil}, {Plunkett}, {Prochaska}, {Rastogi}, {Reddy Janga},
  {Sabater}, {Sakurikar}, {Seifert}, {Sherbert}, {Sherwood-Taylor}, {Shih},
  {Sick}, {Silbiger}, {Singanamalla}, {Singer}, {Sladen}, {Sooley},
  {Sornarajah}, {Streicher}, {Teuben}, {Thomas}, {Tremblay}, {Turner},
  {Terr{\'o}n}, {van Kerkwijk}, {de la Vega}, {Watkins}, {Weaver}, {Whitmore},
  {Woillez}, {Zabalza}, \& {Astropy Contributors}}]{astropy18}
{Astropy Collaboration}, {Price-Whelan}, A.~M., {Sip{\H{o}}cz}, B.~M., {et~al.}
  2018, \aj, 156, 123

\bibitem[{{Astropy Collaboration} {et~al.}(2013){Astropy Collaboration},
  {Robitaille}, {Tollerud}, {Greenfield}, {Droettboom}, {Bray}, {Aldcroft},
  {Davis}, {Ginsburg}, {Price-Whelan}, {Kerzendorf}, {Conley}, {Crighton},
  {Barbary}, {Muna}, {Ferguson}, {Grollier}, {Parikh}, {Nair}, {Unther},
  {Deil}, {Woillez}, {Conseil}, {Kramer}, {Turner}, {Singer}, {Fox}, {Weaver},
  {Zabalza}, {Edwards}, {Azalee Bostroem}, {Burke}, {Casey}, {Crawford},
  {Dencheva}, {Ely}, {Jenness}, {Labrie}, {Lim}, {Pierfederici}, {Pontzen},
  {Ptak}, {Refsdal}, {Servillat}, \& {Streicher}}]{astropy13}
{Astropy Collaboration}, {Robitaille}, T.~P., {Tollerud}, E.~J., {et~al.} 2013,
  \aap, 558, A33

\bibitem[{{Ba{\~n}ados} {et~al.}(2021){Ba{\~n}ados}, {Mazzucchelli}, {Momjian},
  {Eilers}, {Wang}, {Schindler}, {Connor}, {Andika}, {Barth}, {Carilli},
  {Davies}, {Decarli}, {Fan}, {Farina}, {Hennawi}, {Pensabene}, {Stern},
  {Venemans}, {Wenzl}, \& {Yang}}]{banados21}
{Ba{\~n}ados}, E., {Mazzucchelli}, C., {Momjian}, E., {et~al.} 2021, \apj, 909,
  80

\bibitem[{{Ba{\~n}ados} {et~al.}(2016){Ba{\~n}ados}, {Venemans}, {Decarli},
  {Farina}, {Mazzucchelli}, {Walter}, {Fan}, {Stern}, {Schlafly}, {Chambers},
  {Rix}, {Jiang}, {McGreer}, {Simcoe}, {Wang}, {Yang}, {Morganson}, {De Rosa},
  {Greiner}, {Balokovi{\'c}}, {Burgett}, {Cooper}, {Draper}, {Flewelling},
  {Hodapp}, {Jun}, {Kaiser}, {Kudritzki}, {Magnier}, {Metcalfe}, {Miller},
  {Schindler}, {Tonry}, {Wainscoat}, {Waters}, \& {Yang}}]{banados16}
{Ba{\~n}ados}, E., {Venemans}, B.~P., {Decarli}, R., {et~al.} 2016, \apjs, 227,
  11

\bibitem[{{Barger} {et~al.}(2005){Barger}, {Cowie}, {Mushotzky}, {Yang},
  {Wang}, {Steffen}, \& {Capak}}]{barger05}
{Barger}, A.~J., {Cowie}, L.~L., {Mushotzky}, R.~F., {et~al.} 2005, \aj, 129,
  578

\bibitem[{{Barlow-Hall} {et~al.}(2022){Barlow-Hall}, {Delaney}, {Aird},
  {Evans}, {Osborne}, \& {Watson}}]{barlow_hall22}
{Barlow-Hall}, C.~L., {Delaney}, J., {Aird}, J., {et~al.} 2022, arXiv e-prints,
  arXiv:2201.11139

\bibitem[{{Belladitta} {et~al.}(2020){Belladitta}, {Moretti}, {Caccianiga},
  {Spingola}, {Severgnini}, {Della Ceca}, {Ghisellini}, {Dallacasa},
  {Sbarrato}, {Cicone}, {Cassar{\`a}}, \& {Pedani}}]{belladitta20}
{Belladitta}, S., {Moretti}, A., {Caccianiga}, A., {et~al.} 2020, \aap, 635, L7

\bibitem[{{Boller} {et~al.}(1996){Boller}, {Brandt}, \& {Fink}}]{boller96}
{Boller}, T., {Brandt}, W.~N., \& {Fink}, H. 1996, \aap, 305, 53

\bibitem[{{Boquien} {et~al.}(2019){Boquien}, {Burgarella}, {Roehlly}, {Buat},
  {Ciesla}, {Corre}, {Inoue}, \& {Salas}}]{boquien19}
{Boquien}, M., {Burgarella}, D., {Roehlly}, Y., {et~al.} 2019, \aap, 622, A103

\bibitem[{{Boroson} \& {Green}(1992)}]{boroson92}
{Boroson}, T.~A. \& {Green}, R.~F. 1992, \apjs, 80, 109

\bibitem[{{Bouwens} {et~al.}(2015){Bouwens}, {Illingworth}, {Oesch}, {Trenti},
  {Labb{\'e}}, {Bradley}, {Carollo}, {van Dokkum}, {Gonzalez}, {Holwerda},
  {Franx}, {Spitler}, {Smit}, \& {Magee}}]{bouwen15}
{Bouwens}, R.~J., {Illingworth}, G.~D., {Oesch}, P.~A., {et~al.} 2015, \apj,
  803, 34

\bibitem[{{Brandt} {et~al.}(1997){Brandt}, {Mathur}, \& {Elvis}}]{brandt97}
{Brandt}, W.~N., {Mathur}, S., \& {Elvis}, M. 1997, \mnras, 285, L25

\bibitem[{{Brandt} {et~al.}(2002){Brandt}, {Schneider}, {Fan}, {Strauss},
  {Gunn}, {Richards}, {Anderson}, {Vanden Berk}, {Bahcall}, {Brinkmann},
  {Brunner}, {Chen}, {Hennessy}, {Lamb}, {Voges}, \& {York}}]{brandt02}
{Brandt}, W.~N., {Schneider}, D.~P., {Fan}, X., {et~al.} 2002, \apjl, 569, L5

\bibitem[{{Brunner} {et~al.}(2022){Brunner}, {Liu}, {Lamer}, {Georgakakis},
  {Merloni}, {Brusa}, {Bulbul}, {Dennerl}, {Friedrich}, {Liu}, {Maitra},
  {Nandra}, {Ramos-Ceja}, {Sanders}, {Stewart}, {Boller}, {Buchner}, {Clerc},
  {Comparat}, {Dwelly}, {Eckert}, {Finoguenov}, {Freyberg}, {Ghirardini},
  {Gueguen}, {Haberl}, {Kreykenbohm}, {Krumpe}, {Osterhage}, {Pacaud},
  {Predehl}, {Reiprich}, {Robrade}, {Salvato}, {Santangelo}, {Schrabback},
  {Schwope}, \& {Wilms}}]{brunner21}
{Brunner}, H., {Liu}, T., {Lamer}, G., {et~al.} 2022, \aap, 661, A1

\bibitem[{{Brusa} {et~al.}(2009){Brusa}, {Comastri}, {Gilli}, {Hasinger},
  {Iwasawa}, {Mainieri}, {Mignoli}, {Salvato}, {Zamorani}, {Bongiorno},
  {Cappelluti}, {Civano}, {Fiore}, {Merloni}, {Silverman}, {Trump}, {Vignali},
  {Capak}, {Elvis}, {Ilbert}, {Impey}, \& {Lilly}}]{brusa08}
{Brusa}, M., {Comastri}, A., {Gilli}, R., {et~al.} 2009, \apj, 693, 8

\bibitem[{{Bruzual} \& {Charlot}(2003)}]{bruzual03}
{Bruzual}, G. \& {Charlot}, S. 2003, \mnras, 344, 1000

\bibitem[{{Buchner}(2021)}]{buchner21}
{Buchner}, J. 2021, The Journal of Open Source Software, 6, 3001

\bibitem[{{Buchner} {et~al.}(2015){Buchner}, {Georgakakis}, {Nandra},
  {Brightman}, {Menzel}, {Liu}, {Hsu}, {Salvato}, {Rangel}, {Aird}, {Merloni},
  \& {Ross}}]{buchner15}
{Buchner}, J., {Georgakakis}, A., {Nandra}, K., {et~al.} 2015, \apj, 802, 89

\bibitem[{{Buchner} {et~al.}(2014){Buchner}, {Georgakakis}, {Nandra}, {Hsu},
  {Rangel}, {Brightman}, {Merloni}, {Salvato}, {Donley}, \&
  {Kocevski}}]{buchner14}
{Buchner}, J., {Georgakakis}, A., {Nandra}, K., {et~al.} 2014, \aap, 564, A125

\bibitem[{{Calzetti} {et~al.}(2000){Calzetti}, {Armus}, {Bohlin}, {Kinney},
  {Koornneef}, \& {Storchi-Bergmann}}]{calzetti00}
{Calzetti}, D., {Armus}, L., {Bohlin}, R.~C., {et~al.} 2000, \apj, 533, 682

\bibitem[{{Carilli} {et~al.}(2010){Carilli}, {Wang}, {Fan}, {Walter}, {Kurk},
  {Riechers}, {Wagg}, {Hennawi}, {Jiang}, {Menten}, {Bertoldi}, {Strauss}, \&
  {Cox}}]{carilli10}
{Carilli}, C.~L., {Wang}, R., {Fan}, X., {et~al.} 2010, \apj, 714, 834

\bibitem[{{Cash}(1979)}]{cash79}
{Cash}, W. 1979, \apj, 228, 939

\bibitem[{{Civano} {et~al.}(2011){Civano}, {Brusa}, {Comastri}, {Elvis},
  {Salvato}, {Zamorani}, {Capak}, {Fiore}, {Gilli}, {Hao}, {Ikeda}, {Kakazu},
  {Kartaltepe}, {Masters}, {Miyaji}, {Mignoli}, {Puccetti}, {Shankar},
  {Silverman}, {Vignali}, {Zezas}, \& {Koekemoer}}]{civano11}
{Civano}, F., {Brusa}, M., {Comastri}, A., {et~al.} 2011, \apj, 741, 91

\bibitem[{{Collier} {et~al.}(2001){Collier}, {Crenshaw}, {Peterson}, {Brandt},
  {Clavel}, {Edelson}, {George}, {Horne}, {Kriss}, {Mathur}, {Netzer},
  {O'Brien}, {Pogge}, {Pounds}, {Romano}, {Shemmer}, {Turner}, \&
  {Wamsteker}}]{collier01}
{Collier}, S., {Crenshaw}, D.~M., {Peterson}, B.~M., {et~al.} 2001, \apj, 561,
  146

\bibitem[{{Collin} \& {Kawaguchi}(2004)}]{collin04}
{Collin}, S. \& {Kawaguchi}, T. 2004, \aap, 426, 797

\bibitem[{{Crummy} {et~al.}(2006){Crummy}, {Fabian}, {Gallo}, \&
  {Ross}}]{crummy06}
{Crummy}, J., {Fabian}, A.~C., {Gallo}, L., \& {Ross}, R.~R. 2006, \mnras, 365,
  1067

\bibitem[{{Davies} {et~al.}(2019){Davies}, {Hennawi}, \& {Eilers}}]{davies19}
{Davies}, F.~B., {Hennawi}, J.~F., \& {Eilers}, A.-C. 2019, \apjl, 884, L19

\bibitem[{{Davies} {et~al.}(2020){Davies}, {Hennawi}, \& {Eilers}}]{davies20}
{Davies}, F.~B., {Hennawi}, J.~F., \& {Eilers}, A.-C. 2020, \mnras, 493, 1330

\bibitem[{{Delvecchio} {et~al.}(2014){Delvecchio}, {Gruppioni}, {Pozzi},
  {Berta}, {Zamorani}, {Cimatti}, {Lutz}, {Scott}, {Vignali}, {Cresci},
  {Feltre}, {Cooray}, {Vaccari}, {Fritz}, {Le Floc'h}, {Magnelli}, {Popesso},
  {Oliver}, {Bock}, {Carollo}, {Contini}, {Le F{\'e}vre}, {Lilly}, {Mainieri},
  {Renzini}, \& {Scodeggio}}]{delvechhio14}
{Delvecchio}, I., {Gruppioni}, C., {Pozzi}, F., {et~al.} 2014, \mnras, 439,
  2736

\bibitem[{{Dey} {et~al.}(2019){Dey}, {Schlegel}, {Lang}, {Blum}, {Burleigh},
  {Fan}, {Findlay}, {Finkbeiner}, {Herrera}, {Juneau}, {Landriau}, {Levi},
  {McGreer}, {Meisner}, {Myers}, {Moustakas}, {Nugent}, {Patej}, {Schlafly},
  {Walker}, {Valdes}, {Weaver}, {Y{\`e}che}, {Zou}, {Zhou}, {Abareshi},
  {Abbott}, {Abolfathi}, {Aguilera}, {Alam}, {Allen}, {Alvarez}, {Annis},
  {Ansarinejad}, {Aubert}, {Beechert}, {Bell}, {BenZvi}, {Beutler}, {Bielby},
  {Bolton}, {Brice{\~n}o}, {Buckley-Geer}, {Butler}, {Calamida}, {Carlberg},
  {Carter}, {Casas}, {Castander}, {Choi}, {Comparat}, {Cukanovaite}, {Delubac},
  {DeVries}, {Dey}, {Dhungana}, {Dickinson}, {Ding}, {Donaldson}, {Duan},
  {Duckworth}, {Eftekharzadeh}, {Eisenstein}, {Etourneau}, {Fagrelius},
  {Farihi}, {Fitzpatrick}, {Font-Ribera}, {Fulmer}, {G{\"a}nsicke},
  {Gaztanaga}, {George}, {Gerdes}, {Gontcho}, {Gorgoni}, {Green}, {Guy},
  {Harmer}, {Hernandez}, {Honscheid}, {Huang}, {James}, {Jannuzi}, {Jiang},
  {Joyce}, {Karcher}, {Karkar}, {Kehoe}, {Kneib}, {Kueter-Young}, {Lan},
  {Lauer}, {Le Guillou}, {Le Van Suu}, {Lee}, {Lesser}, {Perreault Levasseur},
  {Li}, {Mann}, {Marshall}, {Mart{\'\i}nez-V{\'a}zquez}, {Martini}, {du Mas des
  Bourboux}, {McManus}, {Meier}, {M{\'e}nard}, {Metcalfe},
  {Mu{\~n}oz-Guti{\'e}rrez}, {Najita}, {Napier}, {Narayan}, {Newman}, {Nie},
  {Nord}, {Norman}, {Olsen}, {Paat}, {Palanque-Delabrouille}, {Peng},
  {Poppett}, {Poremba}, {Prakash}, {Rabinowitz}, {Raichoor}, {Rezaie},
  {Robertson}, {Roe}, {Ross}, {Ross}, {Rudnick}, {Safonova}, {Saha},
  {S{\'a}nchez}, {Savary}, {Schweiker}, {Scott}, {Seo}, {Shan}, {Silva},
  {Slepian}, {Soto}, {Sprayberry}, {Staten}, {Stillman}, {Stupak}, {Summers},
  {Sien Tie}, {Tirado}, {Vargas-Maga{\~n}a}, {Vivas}, {Wechsler}, {Williams},
  {Yang}, {Yang}, {Yapici}, {Zaritsky}, {Zenteno}, {Zhang}, {Zhang}, {Zhou}, \&
  {Zhou}}]{dey19}
{Dey}, A., {Schlegel}, D.~J., {Lang}, D., {et~al.} 2019, \aj, 157, 168

\bibitem[{{Dubois} {et~al.}(2015){Dubois}, {Volonteri}, {Silk}, {Devriendt},
  {Slyz}, \& {Teyssier}}]{dubois15}
{Dubois}, Y., {Volonteri}, M., {Silk}, J., {et~al.} 2015, \mnras, 452, 1502

\bibitem[{{Duras} {et~al.}(2017){Duras}, {Bongiorno}, {Piconcelli}, {Bianchi},
  {Pappalardo}, {Valiante}, {Bischetti}, {Feruglio}, {Martocchia}, {Schneider},
  {Vietri}, {Vignali}, {Zappacosta}, {La Franca}, \& {Fiore}}]{duras17}
{Duras}, F., {Bongiorno}, A., {Piconcelli}, E., {et~al.} 2017, \aap, 604, A67

\bibitem[{{Duras} {et~al.}(2020){Duras}, {Bongiorno}, {Ricci}, {Piconcelli},
  {Shankar}, {Lusso}, {Bianchi}, {Fiore}, {Maiolino}, {Marconi}, {Onori},
  {Sani}, {Schneider}, {Vignali}, \& {La Franca}}]{duras20}
{Duras}, F., {Bongiorno}, A., {Ricci}, F., {et~al.} 2020, \aap, 636, A73

\bibitem[{{Ebizuka} {et~al.}(2011){Ebizuka}, {Ichiyama}, {Yamada}, {Tokoku},
  {Onodera}, {Hanesaka}, {Kodate}, {Katsuno Uchimoto}, {Maruyama}, {Shimasaku},
  {Tanaka}, {Yoshikawa}, {Kashikawa}, {Iye}, \& {Ichikawa}}]{ebizuka11}
{Ebizuka}, N., {Ichiyama}, K., {Yamada}, T., {et~al.} 2011, \pasj, 63, 605

\bibitem[{{Eilers} {et~al.}(2017){Eilers}, {Davies}, {Hennawi}, {Prochaska},
  {Luki{\'c}}, \& {Mazzucchelli}}]{eilers17}
{Eilers}, A.-C., {Davies}, F.~B., {Hennawi}, J.~F., {et~al.} 2017, \apj, 840,
  24

\bibitem[{{Eilers} {et~al.}(2021){Eilers}, {Hennawi}, {Davies}, \&
  {Simcoe}}]{eilers21}
{Eilers}, A.-C., {Hennawi}, J.~F., {Davies}, F.~B., \& {Simcoe}, R.~A. 2021,
  \apj, 917, 38

\bibitem[{{Eilers} {et~al.}(2020){Eilers}, {Hennawi}, {Decarli}, {Davies},
  {Venemans}, {Walter}, {Ba{\~n}ados}, {Fan}, {Farina}, {Mazzucchelli},
  {Novak}, {Schindler}, {Simcoe}, {Wang}, \& {Yang}}]{eilers20}
{Eilers}, A.-C., {Hennawi}, J.~F., {Decarli}, R., {et~al.} 2020, \apj, 900, 37

\bibitem[{{Fabian} \& {Iwasawa}(1999)}]{fabian99}
{Fabian}, A.~C. \& {Iwasawa}, K. 1999, \mnras, 303, L34

\bibitem[{{Fan} {et~al.}(2001){Fan}, {Narayanan}, {Lupton}, {Strauss}, {Knapp},
  {Becker}, {White}, {Pentericci}, {Leggett}, {Haiman}, {Gunn}, {Ivezi{\'c}},
  {Schneider}, {Anderson}, {Brinkmann}, {Bahcall}, {Connolly}, {Csabai}, {Doi},
  {Fukugita}, {Geballe}, {Grebel}, {Harbeck}, {Hennessy}, {Lamb}, {Miknaitis},
  {Munn}, {Nichol}, {Okamura}, {Pier}, {Prada}, {Richards}, {Szalay}, \&
  {York}}]{fan01}
{Fan}, X., {Narayanan}, V.~K., {Lupton}, R.~H., {et~al.} 2001, \aj, 122, 2833

\bibitem[{{Fan} {et~al.}(2006){Fan}, {Strauss}, {Becker}, {White}, {Gunn},
  {Knapp}, {Richards}, {Schneider}, {Brinkmann}, \& {Fukugita}}]{fan06}
{Fan}, X., {Strauss}, M.~A., {Becker}, R.~H., {et~al.} 2006, \aj, 132, 117

\bibitem[{{Feltre} {et~al.}(2012){Feltre}, {Hatziminaoglou}, {Fritz}, \&
  {Franceschini}}]{feltre12}
{Feltre}, A., {Hatziminaoglou}, E., {Fritz}, J., \& {Franceschini}, A. 2012,
  \mnras, 426, 120

\bibitem[{Foreman-Mackey(2016)}]{corner}
Foreman-Mackey, D. 2016, The Journal of Open Source Software, 1, 24

\bibitem[{{Fotopoulou} {et~al.}(2016){Fotopoulou}, {Buchner},
  {Georgantopoulos}, {Hasinger}, {Salvato}, {Georgakakis}, {Cappelluti},
  {Ranalli}, {Hsu}, {Brusa}, {Comastri}, {Miyaji}, {Nandra}, {Aird}, \&
  {Paltani}}]{fotopoulo16}
{Fotopoulou}, S., {Buchner}, J., {Georgantopoulos}, I., {et~al.} 2016, \aap,
  587, A142

\bibitem[{{Francis} {et~al.}(1992){Francis}, {Hewett}, {Foltz}, \&
  {Chaffee}}]{francis92}
{Francis}, P.~J., {Hewett}, P.~C., {Foltz}, C.~B., \& {Chaffee}, F.~H. 1992,
  \apj, 398, 476

\bibitem[{{Georgakakis} {et~al.}(2015){Georgakakis}, {Aird}, {Buchner},
  {Salvato}, {Menzel}, {Brandt}, {McGreer}, {Dwelly}, {Mountrichas}, {Koki},
  {Georgantopoulos}, {Hsu}, {Merloni}, {Liu}, {Nandra}, \&
  {Ross}}]{georgakakis15}
{Georgakakis}, A., {Aird}, J., {Buchner}, J., {et~al.} 2015, \mnras, 453, 1946

\bibitem[{{Georgakakis} {et~al.}(2008){Georgakakis}, {Nandra}, {Laird}, {Aird},
  \& {Trichas}}]{georgakakis08}
{Georgakakis}, A., {Nandra}, K., {Laird}, E.~S., {Aird}, J., \& {Trichas}, M.
  2008, \mnras, 388, 1205

\bibitem[{{Gierli{\'n}ski} \& {Done}(2004)}]{gierlinski04}
{Gierli{\'n}ski}, M. \& {Done}, C. 2004, \mnras, 349, L7

\bibitem[{{Goodrich}(1989)}]{goodrich89}
{Goodrich}, R.~W. 1989, \apj, 342, 224

\bibitem[{{Grupe} {et~al.}(2010){Grupe}, {Komossa}, {Leighly}, \&
  {Page}}]{grupe10}
{Grupe}, D., {Komossa}, S., {Leighly}, K.~M., \& {Page}, K.~L. 2010, \apjs,
  187, 64

\bibitem[{{Habouzit} {et~al.}(2017){Habouzit}, {Volonteri}, \&
  {Dubois}}]{habouzit17}
{Habouzit}, M., {Volonteri}, M., \& {Dubois}, Y. 2017, \mnras, 468, 3935

\bibitem[{{Haiman}(2013)}]{haiman13}
{Haiman}, Z. 2013, in Astrophysics and Space Science Library, Vol. 396, The
  First Galaxies, ed. T.~{Wiklind}, B.~{Mobasher}, \& V.~{Bromm}, 293

\bibitem[{{Haiman} \& {Cen}(2001)}]{haiman01}
{Haiman}, Z. \& {Cen}, R. 2001, in Astronomical Society of the Pacific
  Conference Series, Vol. 222, The Physics of Galaxy Formation, ed.
  M.~{Umemura} \& H.~{Susa}, 101

\bibitem[{{Hasinger} {et~al.}(2005){Hasinger}, {Miyaji}, \&
  {Schmidt}}]{hasinger05}
{Hasinger}, G., {Miyaji}, T., \& {Schmidt}, M. 2005, \aap, 441, 417

\bibitem[{{HI4PI Collaboration} {et~al.}(2016){HI4PI Collaboration}, {Ben
  Bekhti}, {Fl{\"o}er}, {Keller}, {Kerp}, {Lenz}, {Winkel}, {Bailin},
  {Calabretta}, {Dedes}, {Ford}, {Gibson}, {Haud}, {Janowiecki}, {Kalberla},
  {Lockman}, {McClure-Griffiths}, {Murphy}, {Nakanishi}, {Pisano}, \&
  {Staveley-Smith}}]{hi4pi16}
{HI4PI Collaboration}, {Ben Bekhti}, N., {Fl{\"o}er}, L., {et~al.} 2016, \aap,
  594, A116

\bibitem[{{Ichikawa} {et~al.}(2006){Ichikawa}, {Suzuki}, {Tokoku}, {Uchimoto},
  {Konishi}, {Yoshikawa}, {Yamada}, {Tanaka}, {Omata}, \&
  {Nishimura}}]{ichikawa06}
{Ichikawa}, T., {Suzuki}, R., {Tokoku}, C., {et~al.} 2006, in Society of
  Photo-Optical Instrumentation Engineers (SPIE) Conference Series, Vol. 6269,
  Society of Photo-Optical Instrumentation Engineers (SPIE) Conference Series,
  ed. I.~S. {McLean} \& M.~{Iye}, 626916

\bibitem[{{Ishimoto} {et~al.}(2020){Ishimoto}, {Kashikawa}, {Onoue},
  {Matsuoka}, {Izumi}, {Strauss}, {Fujimoto}, {Imanishi}, {Ito}, {Iwasawa},
  {Kawaguchi}, {Lee}, {Liang}, {Lu}, {Momose}, {Toba}, \&
  {Uchiyama}}]{ichimoto20}
{Ishimoto}, R., {Kashikawa}, N., {Onoue}, M., {et~al.} 2020, \apj, 903, 60

\bibitem[{Jiang {et~al.}(2016)Jiang, McGreer, Fan, Strauss, Bañados, Becker,
  Bian, Farnsworth, Shen, Wang, Wang, Wang, White, Wu, Wu, Yang, \&
  Yang}]{jiang16}
Jiang, L., McGreer, I.~D., Fan, X., {et~al.} 2016, The Astrophysical Journal,
  833, 222

\bibitem[{Johnson \& Haardt(2016)}]{johnson16}
Johnson, J.~L. \& Haardt, F. 2016, Publications of the Astronomical Society of
  Australia, 33, e007

\bibitem[{{Just} {et~al.}(2007){Just}, {Brandt}, {Shemmer}, {Steffen},
  {Schneider}, {Chartas}, \& {Garmire}}]{just07}
{Just}, D.~W., {Brandt}, W.~N., {Shemmer}, O., {et~al.} 2007, \apj, 665, 1004

\bibitem[{{Kara} {et~al.}(2017){Kara}, {Garc{\'\i}a}, {Lohfink}, {Fabian},
  {Reynolds}, {Tombesi}, \& {Wilkins}}]{kara17}
{Kara}, E., {Garc{\'\i}a}, J.~A., {Lohfink}, A., {et~al.} 2017, \mnras, 468,
  3489

\bibitem[{{Kellermann} {et~al.}(1989){Kellermann}, {Sramek}, {Schmidt},
  {Shaffer}, \& {Green}}]{kellerman89}
{Kellermann}, K.~I., {Sramek}, R., {Schmidt}, M., {Shaffer}, D.~B., \& {Green},
  R. 1989, \aj, 98, 1195

\bibitem[{{Khorunzhev} {et~al.}(2021){Khorunzhev}, {Meshcheryakov}, {Medvedev},
  {Borisov}, {Burenin}, {Krivonos}, {Uklein}, {Shablovinskaya}, {Afanasiev},
  {Dodonov}, {Sunyaev}, {Sazonov}, \& {Gilfanov}}]{khorunzev21}
{Khorunzhev}, G.~A., {Meshcheryakov}, A.~V., {Medvedev}, P.~S., {et~al.} 2021,
  Astronomy Letters, 47, 123

\bibitem[{{Khrykin} {et~al.}(2016){Khrykin}, {Hennawi}, {McQuinn}, \&
  {Worseck}}]{khrykin16}
{Khrykin}, I.~S., {Hennawi}, J.~F., {McQuinn}, M., \& {Worseck}, G. 2016, \apj,
  824, 133

\bibitem[{{Khrykin} {et~al.}(2021){Khrykin}, {Hennawi}, {Worseck}, \&
  {Davies}}]{khrykin21}
{Khrykin}, I.~S., {Hennawi}, J.~F., {Worseck}, G., \& {Davies}, F.~B. 2021,
  \mnras, 505, 649

\bibitem[{{Koptelova} {et~al.}(2019){Koptelova}, {Hwang}, {Malkan}, \&
  {Yu}}]{koptelova19}
{Koptelova}, E., {Hwang}, C.-Y., {Malkan}, M.~A., \& {Yu}, P.-C. 2019, \apj,
  882, 144

\bibitem[{{Koptelova} {et~al.}(2017){Koptelova}, {Hwang}, {Yu}, {Chen}, \&
  {Guo}}]{koptelova17}
{Koptelova}, E., {Hwang}, C.-Y., {Yu}, P.-C., {Chen}, W.-P., \& {Guo}, J.-K.
  2017, Scientific Reports, 7, 41617

\bibitem[{{Latif} \& {Ferrara}(2016)}]{latif16}
{Latif}, M.~A. \& {Ferrara}, A. 2016, \pasa, 33, e051

\bibitem[{{Lawrence} {et~al.}(2007){Lawrence}, {Warren}, {Almaini}, {Edge},
  {Hambly}, {Jameson}, {Lucas}, {Casali}, {Adamson}, {Dye}, {Emerson},
  {Foucaud}, {Hewett}, {Hirst}, {Hodgkin}, {Irwin}, {Lodieu}, {McMahon},
  {Simpson}, {Smail}, {Mortlock}, \& {Folger}}]{lawrence07}
{Lawrence}, A., {Warren}, S.~J., {Almaini}, O., {et~al.} 2007, \mnras, 379,
  1599

\bibitem[{{Leitherer} {et~al.}(2002){Leitherer}, {Li}, {Calzetti}, \&
  {Heckman}}]{leitherer02}
{Leitherer}, C., {Li}, I.~H., {Calzetti}, D., \& {Heckman}, T.~M. 2002, \apjs,
  140, 303

\bibitem[{{Liu} {et~al.}(2022{\natexlab{a}}){Liu}, {Buchner}, {Nandra},
  {Merloni}, {Dwelly}, {Sanders}, {Salvato}, {Arcodia}, {Brusa}, {Wolf},
  {Georgakakis}, {Boller}, {Krumpe}, {Lamer}, {Waddell}, {Urrutia}, {Schwope},
  {Robrade}, {Wilms}, {Dauser}, {Comparat}, {Toba}, {Ichikawa}, {Iwasawa},
  {Shen}, \& {Medel}}]{liu22}
{Liu}, T., {Buchner}, J., {Nandra}, K., {et~al.} 2022{\natexlab{a}}, \aap, 661,
  A5

\bibitem[{{Liu} {et~al.}(2022{\natexlab{b}}){Liu}, {Merloni}, {Comparat},
  {Nandra}, {Sanders}, {Lamer}, {Buchner}, {Dwelly}, {Freyberg}, {Malyali},
  {Georgakakis}, {Salvato}, {Brunner}, {Brusa}, {Klein}, {Ghirardini}, {Clerc},
  {Pacaud}, {Bulbul}, {Liu}, {Schwope}, {Robrade}, {Wilms}, {Dauser},
  {Ramos-Ceja}, {Reiprich}, {Boller}, \& {Wolf}}]{liu21}
{Liu}, T., {Merloni}, A., {Comparat}, J., {et~al.} 2022{\natexlab{b}}, \aap,
  661, A27

\bibitem[{{Lusso} {et~al.}(2012){Lusso}, {Comastri}, {Simmons}, {Mignoli},
  {Zamorani}, {Vignali}, {Brusa}, {Shankar}, {Lutz}, {Trump}, {Maiolino},
  {Gilli}, {Bolzonella}, {Puccetti}, {Salvato}, {Impey}, {Civano}, {Elvis},
  {Mainieri}, {Silverman}, {Koekemoer}, {Bongiorno}, {Merloni}, {Berta}, {Le
  Floc'h}, {Magnelli}, {Pozzi}, \& {Riguccini}}]{lusso12}
{Lusso}, E., {Comastri}, A., {Simmons}, B.~D., {et~al.} 2012, \mnras, 425, 623

\bibitem[{{Lusso} \& {Risaliti}(2016)}]{lusso16}
{Lusso}, E. \& {Risaliti}, G. 2016, \apj, 819, 154

\bibitem[{{Magdziarz} {et~al.}(1998){Magdziarz}, {Blaes}, {Zdziarski},
  {Johnson}, \& {Smith}}]{magdziarz98}
{Magdziarz}, P., {Blaes}, O.~M., {Zdziarski}, A.~A., {Johnson}, W.~N., \&
  {Smith}, D.~A. 1998, \mnras, 301, 179

\bibitem[{{Marocco} {et~al.}(2021){Marocco}, {Eisenhardt}, {Fowler},
  {Kirkpatrick}, {Meisner}, {Schlafly}, {Stanford}, {Garcia}, {Caselden},
  {Cushing}, {Cutri}, {Faherty}, {Gelino}, {Gonzalez}, {Jarrett}, {Koontz},
  {Mainzer}, {Marchese}, {Mobasher}, {Schlegel}, {Stern}, {Teplitz}, \&
  {Wright}}]{marocco21}
{Marocco}, F., {Eisenhardt}, P. R.~M., {Fowler}, J.~W., {et~al.} 2021, \apjs,
  253, 8

\bibitem[{{Marziani} {et~al.}(2018{\natexlab{a}}){Marziani}, {del Olmo},
  {D'Onofrio}, {Dultzin}, {Negrete}, {Mart{\'\i}nez-Aldama}, {Bon}, {Bon}, \&
  {Stirpe}}]{marziani18b}
{Marziani}, P., {del Olmo}, A., {D'Onofrio}, M., {et~al.} 2018{\natexlab{a}},
  in Revisiting Narrow-Line Seyfert 1 Galaxies and their Place in the Universe,
  2

\bibitem[{{Marziani} {et~al.}(2018{\natexlab{b}}){Marziani}, {Dultzin},
  {Sulentic}, {Del Olmo}, {Negrete}, {Mart{\'\i}nez-Aldama}, {D'Onofrio},
  {Bon}, {Bon}, \& {Stirpe}}]{marziani18}
{Marziani}, P., {Dultzin}, D., {Sulentic}, J.~W., {et~al.} 2018{\natexlab{b}},
  Frontiers in Astronomy and Space Sciences, 5, 6

\bibitem[{{Matsuoka} {et~al.}(2022){Matsuoka}, {Iwasawa}, {Onoue}, {Izumi},
  {Kashikawa}, {Strauss}, {Imanishi}, {Nagao}, {Akiyama}, {Silverman}, {Asami},
  {Bosch}, {Furusawa}, {Goto}, {Gunn}, {Harikane}, {Ikeda}, {Ishimoto},
  {Kawaguchi}, {Kato}, {Kikuta}, {Kohno}, {Komiyama}, {Lee}, {Lupton},
  {Minezaki}, {Miyazaki}, {Murayama}, {Nishizawa}, {Oguri}, {Ono}, {Ouchi},
  {Price}, {Sameshima}, {Sugiyama}, {Tait}, {Takada}, {Takahashi}, {Takata},
  {Tanaka}, {Toba}, {Utsumi}, {Wang}, \& {Yamashita}}]{matsuoka22}
{Matsuoka}, Y., {Iwasawa}, K., {Onoue}, M., {et~al.} 2022, \apjs, 259, 18

\bibitem[{{Matsuoka} {et~al.}(2019){Matsuoka}, {Iwasawa}, {Onoue}, {Kashikawa},
  {Strauss}, {Lee}, {Imanishi}, {Nagao}, {Akiyama}, {Asami}, {Bosch},
  {Furusawa}, {Goto}, {Gunn}, {Harikane}, {Ikeda}, {Izumi}, {Kawaguchi},
  {Kato}, {Kikuta}, {Kohno}, {Komiyama}, {Koyama}, {Lupton}, {Minezaki},
  {Miyazaki}, {Murayama}, {Niida}, {Nishizawa}, {Noboriguchi}, {Oguri}, {Ono},
  {Ouchi}, {Price}, {Sameshima}, {Schulze}, {Silverman}, {Sugiyama}, {Tait},
  {Takada}, {Takata}, {Tanaka}, {Tang}, {Toba}, {Utsumi}, {Wang}, \&
  {Yamashita}}]{matsuoka19}
{Matsuoka}, Y., {Iwasawa}, K., {Onoue}, M., {et~al.} 2019, \apj, 883, 183

\bibitem[{{Matsuoka} {et~al.}(2018{\natexlab{a}}){Matsuoka}, {Iwasawa},
  {Onoue}, {Kashikawa}, {Strauss}, {Lee}, {Imanishi}, {Nagao}, {Akiyama},
  {Asami}, {Bosch}, {Furusawa}, {Goto}, {Gunn}, {Harikane}, {Ikeda}, {Izumi},
  {Kawaguchi}, {Kato}, {Kikuta}, {Kohno}, {Komiyama}, {Lupton}, {Minezaki},
  {Miyazaki}, {Morokuma}, {Murayama}, {Niida}, {Nishizawa}, {Oguri}, {Ono},
  {Ouchi}, {Price}, {Sameshima}, {Schulze}, {Shirakata}, {Silverman},
  {Sugiyama}, {Tait}, {Takada}, {Takata}, {Tanaka}, {Tang}, {Toba}, {Utsumi},
  {Wang}, \& {Yamashita}}]{matsuoka18}
{Matsuoka}, Y., {Iwasawa}, K., {Onoue}, M., {et~al.} 2018{\natexlab{a}}, \apjs,
  237, 5

\bibitem[{{Matsuoka} {et~al.}(2018{\natexlab{b}}){Matsuoka}, {Onoue},
  {Kashikawa}, {Iwasawa}, {Strauss}, {Nagao}, {Imanishi}, {Lee}, {Akiyama},
  {Asami}, {Bosch}, {Foucaud}, {Furusawa}, {Goto}, {Gunn}, {Harikane}, {Ikeda},
  {Izumi}, {Kawaguchi}, {Kikuta}, {Kohno}, {Komiyama}, {Lupton}, {Minezaki},
  {Miyazaki}, {Morokuma}, {Murayama}, {Niida}, {Nishizawa}, {Oguri}, {Ono},
  {Ouchi}, {Price}, {Sameshima}, {Schulze}, {Shirakata}, {Silverman},
  {Sugiyama}, {Tait}, {Takada}, {Takata}, {Tanaka}, {Tang}, {Toba}, {Utsumi},
  \& {Wang}}]{matsuoka18b}
{Matsuoka}, Y., {Onoue}, M., {Kashikawa}, N., {et~al.} 2018{\natexlab{b}},
  \pasj, 70, S35

\bibitem[{{Matsuoka} {et~al.}(2016){Matsuoka}, {Onoue}, {Kashikawa}, {Iwasawa},
  {Strauss}, {Nagao}, {Imanishi}, {Niida}, {Toba}, {Akiyama}, {Asami}, {Bosch},
  {Foucaud}, {Furusawa}, {Goto}, {Gunn}, {Harikane}, {Ikeda}, {Kawaguchi},
  {Kikuta}, {Komiyama}, {Lupton}, {Minezaki}, {Miyazaki}, {Morokuma},
  {Murayama}, {Nishizawa}, {Ono}, {Ouchi}, {Price}, {Sameshima}, {Silverman},
  {Sugiyama}, {Tait}, {Takada}, {Takata}, {Tanaka}, {Tang}, \&
  {Utsumi}}]{matsuoka16}
{Matsuoka}, Y., {Onoue}, M., {Kashikawa}, N., {et~al.} 2016, \apj, 828, 26

\bibitem[{{Mazzucchelli} {et~al.}(2017){Mazzucchelli}, {Ba{\~n}ados},
  {Venemans}, {Decarli}, {Farina}, {Walter}, {Eilers}, {Rix}, {Simcoe},
  {Stern}, {Fan}, {Schlafly}, {De Rosa}, {Hennawi}, {Chambers}, {Greiner},
  {Burgett}, {Draper}, {Kaiser}, {Kudritzki}, {Magnier}, {Metcalfe}, {Waters},
  \& {Wainscoat}}]{mazzuchelli17}
{Mazzucchelli}, C., {Ba{\~n}ados}, E., {Venemans}, B.~P., {et~al.} 2017, \apj,
  849, 91

\bibitem[{{Medvedev} {et~al.}(2021){Medvedev}, {Gilfanov}, {Sazonov},
  {Schartel}, \& {Sunyaev}}]{medvedev21}
{Medvedev}, P., {Gilfanov}, M., {Sazonov}, S., {Schartel}, N., \& {Sunyaev}, R.
  2021, \mnras, 504, 576

\bibitem[{{Medvedev} {et~al.}(2020){Medvedev}, {Sazonov}, {Gilfanov},
  {Burenin}, {Khorunzhev}, {Meshcheryakov}, {Sunyaev}, {Bikmaev}, \&
  {Irtuganov}}]{medvedev20}
{Medvedev}, P., {Sazonov}, S., {Gilfanov}, M., {et~al.} 2020, \mnras, 497, 1842

\bibitem[{{Merloni} \& {Heinz}(2008)}]{merloni08}
{Merloni}, A. \& {Heinz}, S. 2008, \mnras, 388, 1011

\bibitem[{{Merloni} {et~al.}(2012){Merloni}, {Predehl}, {Becker},
  {B{\"o}hringer}, {Boller}, {Brunner}, {Brusa}, {Dennerl}, {Freyberg},
  {Friedrich}, {Georgakakis}, {Haberl}, {Hasinger}, {Meidinger}, {Mohr},
  {Nandra}, {Rau}, {Reiprich}, {Robrade}, {Salvato}, {Santangelo}, {Sasaki},
  {Schwope}, {Wilms}, \& {German eROSITA Consortium}}]{merloni12}
{Merloni}, A., {Predehl}, P., {Becker}, W., {et~al.} 2012, arXiv e-prints,
  arXiv:1209.3114

\bibitem[{{Miyaji} {et~al.}(2015){Miyaji}, {Hasinger}, {Salvato}, {Brusa},
  {Cappelluti}, {Civano}, {Puccetti}, {Elvis}, {Brunner}, {Fotopoulou}, {Ueda},
  {Griffiths}, {Koekemoer}, {Akiyama}, {Comastri}, {Gilli}, {Lanzuisi},
  {Merloni}, \& {Vignali}}]{miyaji15}
{Miyaji}, T., {Hasinger}, G., {Salvato}, M., {et~al.} 2015, \apj, 804, 104

\bibitem[{Mortlock {et~al.}(2012)Mortlock, Patel, Warren, Hewett, Venemans,
  McMahon, \& Simpson}]{mortlock12}
Mortlock, D.~J., Patel, M., Warren, S.~J., {et~al.} 2012, Monthly Notices of
  the Royal Astronomical Society, 419, 390, publisher: Oxford Academic

\bibitem[{{Nandra} \& {Pounds}(1994)}]{nandra94}
{Nandra}, K. \& {Pounds}, K.~A. 1994, \mnras, 268, 405

\bibitem[{{Nanni} {et~al.}(2018){Nanni}, {Gilli}, {Vignali}, {Mignoli},
  {Comastri}, {Vanzella}, {Zamorani}, {Calura}, {Lanzuisi}, {Brusa}, {Tozzi},
  {Iwasawa}, {Cappi}, {Vito}, {Balmaverde}, {Costa}, {Risaliti}, {Paolillo},
  {Prandoni}, {Liuzzo}, {Rosati}, {Chiaberge}, {Caminha}, {Sani}, {Cappelluti},
  \& {Norman}}]{nanni18}
{Nanni}, R., {Gilli}, R., {Vignali}, C., {et~al.} 2018, \aap, 614, A121

\bibitem[{{Nanni} {et~al.}(2017){Nanni}, {Vignali}, {Gilli}, {Moretti}, \&
  {Brandt}}]{nanni17}
{Nanni}, R., {Vignali}, C., {Gilli}, R., {Moretti}, A., \& {Brandt}, W.~N.
  2017, \aap, 603, A128

\bibitem[{{Ni} {et~al.}(2022){Ni}, {Di Matteo}, {Bird}, {Croft}, {Feng},
  {Chen}, {Tremmel}, {DeGraf}, \& {Li}}]{yueying22}
{Ni}, Y., {Di Matteo}, T., {Bird}, S., {et~al.} 2022, \mnras, 513, 670

\bibitem[{{Ojha} {et~al.}(2020){Ojha}, {Chand}, {Dewangan}, \&
  {Rakshit}}]{ohja20}
{Ojha}, V., {Chand}, H., {Dewangan}, G.~C., \& {Rakshit}, S. 2020, \apj, 896,
  95

\bibitem[{{Onoue} {et~al.}(2019){Onoue}, {Kashikawa}, {Matsuoka}, {Kato},
  {Izumi}, {Nagao}, {Strauss}, {Harikane}, {Imanishi}, {Ito}, {Iwasawa},
  {Kawaguchi}, {Lee}, {Noboriguchi}, {Suh}, {Tanaka}, \& {Toba}}]{onoue19}
{Onoue}, M., {Kashikawa}, N., {Matsuoka}, Y., {et~al.} 2019, \apj, 880, 77

\bibitem[{{Osterbrock} \& {Dahari}(1983)}]{osterbrock83}
{Osterbrock}, D.~E. \& {Dahari}, O. 1983, \apj, 273, 478

\bibitem[{{Osterbrock} \& {Pogge}(1985)}]{osterbrock85}
{Osterbrock}, D.~E. \& {Pogge}, R.~W. 1985, \apj, 297, 166

\bibitem[{{P{\^a}ris} {et~al.}(2011){P{\^a}ris}, {Petitjean}, {Rollinde},
  {Aubourg}, {Busca}, {Charlassier}, {Delubac}, {Hamilton}, {Le Goff},
  {Palanque-Delabrouille}, {Peirani}, {Pichon}, {Rich}, {Vargas-Maga{\~n}a}, \&
  {Y{\`e}che}}]{paris11}
{P{\^a}ris}, I., {Petitjean}, P., {Rollinde}, E., {et~al.} 2011, \aap, 530, A50

\bibitem[{{Petrucci} {et~al.}(2001){Petrucci}, {Haardt}, {Maraschi}, {Grandi},
  {Malzac}, {Matt}, {Nicastro}, {Piro}, {Perola}, \& {De Rosa}}]{petrucci01}
{Petrucci}, P.~O., {Haardt}, F., {Maraschi}, L., {et~al.} 2001, \apj, 556, 716

\bibitem[{{Planck Collaboration} {et~al.}(2020){Planck Collaboration},
  {Aghanim}, {Akrami}, {Ashdown}, {Aumont}, {Baccigalupi}, {Ballardini},
  {Banday}, {Barreiro}, {Bartolo}, {Basak}, {Battye}, {Benabed}, {Bernard},
  {Bersanelli}, {Bielewicz}, {Bock}, {Bond}, {Borrill}, {Bouchet}, {Boulanger},
  {Bucher}, {Burigana}, {Butler}, {Calabrese}, {Cardoso}, {Carron},
  {Challinor}, {Chiang}, {Chluba}, {Colombo}, {Combet}, {Contreras}, {Crill},
  {Cuttaia}, {de Bernardis}, {de Zotti}, {Delabrouille}, {Delouis}, {Di
  Valentino}, {Diego}, {Dor{\'e}}, {Douspis}, {Ducout}, {Dupac}, {Dusini},
  {Efstathiou}, {Elsner}, {En{\ss}lin}, {Eriksen}, {Fantaye}, {Farhang},
  {Fergusson}, {Fernandez-Cobos}, {Finelli}, {Forastieri}, {Frailis},
  {Fraisse}, {Franceschi}, {Frolov}, {Galeotta}, {Galli}, {Ganga},
  {G{\'e}nova-Santos}, {Gerbino}, {Ghosh}, {Gonz{\'a}lez-Nuevo}, {G{\'o}rski},
  {Gratton}, {Gruppuso}, {Gudmundsson}, {Hamann}, {Handley}, {Hansen},
  {Herranz}, {Hildebrandt}, {Hivon}, {Huang}, {Jaffe}, {Jones}, {Karakci},
  {Keih{\"a}nen}, {Keskitalo}, {Kiiveri}, {Kim}, {Kisner}, {Knox},
  {Krachmalnicoff}, {Kunz}, {Kurki-Suonio}, {Lagache}, {Lamarre}, {Lasenby},
  {Lattanzi}, {Lawrence}, {Le Jeune}, {Lemos}, {Lesgourgues}, {Levrier},
  {Lewis}, {Liguori}, {Lilje}, {Lilley}, {Lindholm}, {L{\'o}pez-Caniego},
  {Lubin}, {Ma}, {Mac{\'\i}as-P{\'e}rez}, {Maggio}, {Maino}, {Mandolesi},
  {Mangilli}, {Marcos-Caballero}, {Maris}, {Martin}, {Martinelli},
  {Mart{\'\i}nez-Gonz{\'a}lez}, {Matarrese}, {Mauri}, {McEwen}, {Meinhold},
  {Melchiorri}, {Mennella}, {Migliaccio}, {Millea}, {Mitra},
  {Miville-Desch{\^e}nes}, {Molinari}, {Montier}, {Morgante}, {Moss}, {Natoli},
  {N{\o}rgaard-Nielsen}, {Pagano}, {Paoletti}, {Partridge}, {Patanchon},
  {Peiris}, {Perrotta}, {Pettorino}, {Piacentini}, {Polastri}, {Polenta},
  {Puget}, {Rachen}, {Reinecke}, {Remazeilles}, {Renzi}, {Rocha}, {Rosset},
  {Roudier}, {Rubi{\~n}o-Mart{\'\i}n}, {Ruiz-Granados}, {Salvati}, {Sandri},
  {Savelainen}, {Scott}, {Shellard}, {Sirignano}, {Sirri}, {Spencer},
  {Sunyaev}, {Suur-Uski}, {Tauber}, {Tavagnacco}, {Tenti}, {Toffolatti},
  {Tomasi}, {Trombetti}, {Valenziano}, {Valiviita}, {Van Tent}, {Vibert},
  {Vielva}, {Villa}, {Vittorio}, {Wandelt}, {Wehus}, {White}, {White},
  {Zacchei}, \& {Zonca}}]{planck18}
{Planck Collaboration}, {Aghanim}, N., {Akrami}, Y., {et~al.} 2020, \aap, 641,
  A6

\bibitem[{{Pons} {et~al.}(2020){Pons}, {McMahon}, {Banerji}, \&
  {Reed}}]{pons20}
{Pons}, E., {McMahon}, R.~G., {Banerji}, M., \& {Reed}, S.~L. 2020, \mnras,
  491, 3884

\bibitem[{{Pounds} {et~al.}(1995){Pounds}, {Done}, \& {Osborne}}]{pounds95}
{Pounds}, K.~A., {Done}, C., \& {Osborne}, J.~P. 1995, \mnras, 277, L5

\bibitem[{{Predehl} {et~al.}(2021){Predehl}, {Andritschke}, {Arefiev},
  {Babyshkin}, {Batanov}, {Becker}, {B{\"o}hringer}, {Bogomolov}, {Boller},
  {Borm}, {Bornemann}, {Br{\"a}uninger}, {Br{\"u}ggen}, {Brunner}, {Brusa},
  {Bulbul}, {Buntov}, {Burwitz}, {Burkert}, {Clerc}, {Churazov}, {Coutinho},
  {Dauser}, {Dennerl}, {Doroshenko}, {Eder}, {Emberger}, {Eraerds},
  {Finoguenov}, {Freyberg}, {Friedrich}, {Friedrich}, {F{\"u}rmetz},
  {Georgakakis}, {Gilfanov}, {Granato}, {Grossberger}, {Gueguen}, {Gureev},
  {Haberl}, {H{\"a}lker}, {Hartner}, {Hasinger}, {Huber}, {Ji}, {Kienlin},
  {Kink}, {Korotkov}, {Kreykenbohm}, {Lamer}, {Lomakin}, {Lapshov}, {Liu},
  {Maitra}, {Meidinger}, {Menz}, {Merloni}, {Mernik}, {Mican}, {Mohr},
  {M{\"u}ller}, {Nandra}, {Nazarov}, {Pacaud}, {Pavlinsky}, {Perinati},
  {Pfeffermann}, {Pietschner}, {Ramos-Ceja}, {Rau}, {Reiffers}, {Reiprich},
  {Robrade}, {Salvato}, {Sanders}, {Santangelo}, {Sasaki}, {Scheuerle},
  {Schmid}, {Schmitt}, {Schwope}, {Shirshakov}, {Steinmetz}, {Stewart},
  {Str{\"u}der}, {Sunyaev}, {Tenzer}, {Tiedemann}, {Tr{\"u}mper}, {Voron},
  {Weber}, {Wilms}, \& {Yaroshenko}}]{predehl21}
{Predehl}, P., {Andritschke}, R., {Arefiev}, V., {et~al.} 2021, \aap, 647, A1

\bibitem[{{Rakshit} {et~al.}(2017){Rakshit}, {Stalin}, {Chand}, \&
  {Zhang}}]{rakshit17}
{Rakshit}, S., {Stalin}, C.~S., {Chand}, H., \& {Zhang}, X.-G. 2017, \apjs,
  229, 39

\bibitem[{{Rakshit} {et~al.}(2021){Rakshit}, {Stalin}, {Kotilainen}, \&
  {Shin}}]{rakshit21}
{Rakshit}, S., {Stalin}, C.~S., {Kotilainen}, J., \& {Shin}, J. 2021, \apjs,
  253, 28

\bibitem[{{Richards} {et~al.}(2006){Richards}, {Lacy}, {Storrie-Lombardi},
  {Hall}, {Gallagher}, {Hines}, {Fan}, {Papovich}, {Vanden Berk}, {Trammell},
  {Schneider}, {Vestergaard}, {York}, {Jester}, {Anderson}, {Budav{\'a}ri}, \&
  {Szalay}}]{richards06}
{Richards}, G.~T., {Lacy}, M., {Storrie-Lombardi}, L.~J., {et~al.} 2006, \apjs,
  166, 470

\bibitem[{{Romano} {et~al.}(2002){Romano}, {Turner}, {Mathur}, \&
  {George}}]{romano02}
{Romano}, P., {Turner}, T.~J., {Mathur}, S., \& {George}, I.~M. 2002, \apj,
  564, 162

\bibitem[{{Ross} \& {Fabian}(2005)}]{ross05}
{Ross}, R.~R. \& {Fabian}, A.~C. 2005, \mnras, 358, 211

\bibitem[{{Salvato} {et~al.}(2018){Salvato}, {Buchner}, {Budav{\'a}ri},
  {Dwelly}, {Merloni}, {Brusa}, {Rau}, {Fotopoulou}, \& {Nandra}}]{salvato18}
{Salvato}, M., {Buchner}, J., {Budav{\'a}ri}, T., {et~al.} 2018, \mnras, 473,
  4937

\bibitem[{{Salvato} {et~al.}(2022){Salvato}, {Wolf}, {Dwelly}, {Georgakakis},
  {Brusa}, {Merloni}, {Liu}, {Toba}, {Nandra}, {Lamer}, {Buchner}, {Schneider},
  {Freund}, {Rau}, {Schwope}, {Nishizawa}, {Klein}, {Arcodia}, {Comparat},
  {Musiimenta}, {Nagao}, {Brunner}, {Malyali}, {Finoguenov}, {Anderson},
  {Shen}, {Ibarra-Medel}, {Trump}, {Brandt}, {Urry}, {Rivera}, {Krumpe},
  {Urrutia}, {Miyaji}, {Ichikawa}, {Schneider}, {Fresco}, {Boller}, {Haase},
  {Brownstein}, {Lane}, {Bizyaev}, \& {Nitschelm}}]{salvato21}
{Salvato}, M., {Wolf}, J., {Dwelly}, T., {et~al.} 2022, \aap, 661, A3

\bibitem[{{Seppi} {et~al.}(2022){Seppi}, {Comparat}, {Bulbul}, {Nandra},
  {Merloni}, {Clerc}, {Liu}, {Ghirardini}, {Liu}, {Salvato}, {Sanders},
  {Wilms}, {Dwelly}, {Dauser}, {K{\"o}nig}, {Ramos-Ceja}, {Garrel}, \&
  {Reiprich}}]{seppi22}
{Seppi}, R., {Comparat}, J., {Bulbul}, E., {et~al.} 2022, \aap, 665, A78

\bibitem[{{Shemmer} {et~al.}(2006){Shemmer}, {Brandt}, {Netzer}, {Maiolino}, \&
  {Kaspi}}]{shemmer06}
{Shemmer}, O., {Brandt}, W.~N., {Netzer}, H., {Maiolino}, R., \& {Kaspi}, S.
  2006, \apjl, 646, L29

\bibitem[{{Sijacki} {et~al.}(2015){Sijacki}, {Vogelsberger}, {Genel},
  {Springel}, {Torrey}, {Snyder}, {Nelson}, \& {Hernquist}}]{sijacki15}
{Sijacki}, D., {Vogelsberger}, M., {Genel}, S., {et~al.} 2015, \mnras, 452, 575

\bibitem[{{Soltan}(1982)}]{soltan82}
{Soltan}, A. 1982, \mnras, 200, 115

\bibitem[{{Steffen} {et~al.}(2006){Steffen}, {Strateva}, {Brandt}, {Alexander},
  {Koekemoer}, {Lehmer}, {Schneider}, \& {Vignali}}]{steffen06}
{Steffen}, A.~T., {Strateva}, I., {Brandt}, W.~N., {et~al.} 2006, \aj, 131,
  2826

\bibitem[{{Sulentic} {et~al.}(2000){Sulentic}, {Zwitter}, {Marziani}, \&
  {Dultzin-Hacyan}}]{sulentic00}
{Sulentic}, J.~W., {Zwitter}, T., {Marziani}, P., \& {Dultzin-Hacyan}, D. 2000,
  \apjl, 536, L5

\bibitem[{{Sunyaev} {et~al.}(2021){Sunyaev}, {Arefiev}, {Babyshkin},
  {Bogomolov}, {Borisov}, {Buntov}, {Brunner}, {Burenin}, {Churazov},
  {Coutinho}, {Eder}, {Eismont}, {Freyberg}, {Gilfanov}, {Gureyev}, {Hasinger},
  {Khabibullin}, {Kolmykov}, {Komovkin}, {Krivonos}, {Lapshov}, {Levin},
  {Lomakin}, {Lutovinov}, {Medvedev}, {Merloni}, {Mernik}, {Mikhailov},
  {Molodtsov}, {Mzhelsky}, {M{\"u}ller}, {Nandra}, {Nazarov}, {Pavlinsky},
  {Poghodin}, {Predehl}, {Robrade}, {Sazonov}, {Scheuerle}, {Shirshakov},
  {Tkachenko}, \& {Voron}}]{sunyaev21}
{Sunyaev}, R., {Arefiev}, V., {Babyshkin}, V., {et~al.} 2021, \aap, 656, A132

\bibitem[{{Suzuki} {et~al.}(2005){Suzuki}, {Tytler}, {Kirkman}, {O'Meara}, \&
  {Lubin}}]{suzuki05}
{Suzuki}, N., {Tytler}, D., {Kirkman}, D., {O'Meara}, J.~M., \& {Lubin}, D.
  2005, \apj, 618, 592

\bibitem[{{Suzuki} {et~al.}(2008){Suzuki}, {Tokoku}, {Ichikawa}, {Uchimoto},
  {Konishi}, {Yoshikawa}, {Tanaka}, {Yamada}, {Omata}, \&
  {Nishimura}}]{suzuji08}
{Suzuki}, R., {Tokoku}, C., {Ichikawa}, T., {et~al.} 2008, \pasj, 60, 1347

\bibitem[{{Tsuzuki} {et~al.}(2006){Tsuzuki}, {Kawara}, {Yoshii}, {Oyabu},
  {Tanab{\'e}}, \& {Matsuoka}}]{tsuzuki06}
{Tsuzuki}, Y., {Kawara}, K., {Yoshii}, Y., {et~al.} 2006, \apj, 650, 57

\bibitem[{{Turner} {et~al.}(2001){Turner}, {Romano}, {George}, {Edelson},
  {Collier}, {Mathur}, \& {Peterson}}]{turner2001}
{Turner}, T.~J., {Romano}, P., {George}, I.~M., {et~al.} 2001, \apj, 561, 131

\bibitem[{{Ueda} {et~al.}(2014){Ueda}, {Akiyama}, {Hasinger}, {Miyaji}, \&
  {Watson}}]{ueda14}
{Ueda}, Y., {Akiyama}, M., {Hasinger}, G., {Miyaji}, T., \& {Watson}, M.~G.
  2014, \apj, 786, 104

\bibitem[{{Venemans} {et~al.}(2015){Venemans}, {Verdoes Kleijn}, {Mwebaze},
  {Valentijn}, {Ba{\~n}ados}, {Decarli}, {de Jong}, {Findlay}, {Kuijken}, {La
  Barbera}, {McFarland}, {McMahon}, {Napolitano}, {Sikkema}, \&
  {Sutherland}}]{venemans15}
{Venemans}, B.~P., {Verdoes Kleijn}, G.~A., {Mwebaze}, J., {et~al.} 2015,
  \mnras, 453, 2259

\bibitem[{{Vestergaard} \& {Osmer}(2009)}]{vestergaard09}
{Vestergaard}, M. \& {Osmer}, P.~S. 2009, \apj, 699, 800

\bibitem[{{Vignali} {et~al.}(2003){Vignali}, {Brandt}, {Schneider}, {Garmire},
  \& {Kaspi}}]{vignali03}
{Vignali}, C., {Brandt}, W.~N., {Schneider}, D.~P., {Garmire}, G.~P., \&
  {Kaspi}, S. 2003, \aj, 125, 418

\bibitem[{{Vito} {et~al.}(2019){Vito}, {Brandt}, {Bauer}, {Calura}, {Gilli},
  {Luo}, {Shemmer}, {Vignali}, {Zamorani}, {Brusa}, {Civano}, {Comastri}, \&
  {Nanni}}]{vito19}
{Vito}, F., {Brandt}, W.~N., {Bauer}, F.~E., {et~al.} 2019, \aap, 630, A118

\bibitem[{{Vito} {et~al.}(2018){Vito}, {Brandt}, {Yang}, {Gilli}, {Luo},
  {Vignali}, {Xue}, {Comastri}, {Koekemoer}, {Lehmer}, {Liu}, {Paolillo},
  {Ranalli}, {Schneider}, {Shemmer}, {Volonteri}, \& {Wang}}]{vito18}
{Vito}, F., {Brandt}, W.~N., {Yang}, G., {et~al.} 2018, \mnras, 473, 2378

\bibitem[{{Vito} {et~al.}(2016){Vito}, {Gilli}, {Vignali}, {Brandt},
  {Comastri}, {Yang}, {Lehmer}, {Luo}, {Basu-Zych}, {Bauer}, {Cappelluti},
  {Koekemoer}, {Mainieri}, {Paolillo}, {Ranalli}, {Shemmer}, {Trump}, {Wang},
  \& {Xue}}]{vito16}
{Vito}, F., {Gilli}, R., {Vignali}, C., {et~al.} 2016, \mnras, 463, 348

\bibitem[{{Vito} {et~al.}(2014){Vito}, {Gilli}, {Vignali}, {Comastri}, {Brusa},
  {Cappelluti}, \& {Iwasawa}}]{vito14}
{Vito}, F., {Gilli}, R., {Vignali}, C., {et~al.} 2014, \mnras, 445, 3557

\bibitem[{{Volonteri}(2010)}]{volonteri10}
{Volonteri}, M. 2010, \aapr, 18, 279

\bibitem[{{Volonteri} {et~al.}(2016){Volonteri}, {Dubois}, {Pichon}, \&
  {Devriendt}}]{volonteri16}
{Volonteri}, M., {Dubois}, Y., {Pichon}, C., \& {Devriendt}, J. 2016, \mnras,
  460, 2979

\bibitem[{{Volonteri} {et~al.}(2021){Volonteri}, {Habouzit}, \&
  {Colpi}}]{volonteri21}
{Volonteri}, M., {Habouzit}, M., \& {Colpi}, M. 2021, Nature Reviews Physics,
  3, 732

\bibitem[{{Volonteri} \& {Rees}(2005)}]{volonteri05}
{Volonteri}, M. \& {Rees}, M.~J. 2005, \apj, 633, 624

\bibitem[{{Waddell} \& {Gallo}(2020)}]{waddell20}
{Waddell}, S.~G.~H. \& {Gallo}, L.~C. 2020, \mnras, 498, 5207

\bibitem[{{Walton} {et~al.}(2013){Walton}, {Nardini}, {Fabian}, {Gallo}, \&
  {Reis}}]{walton13}
{Walton}, D.~J., {Nardini}, E., {Fabian}, A.~C., {Gallo}, L.~C., \& {Reis},
  R.~C. 2013, \mnras, 428, 2901

\bibitem[{{Wang} {et~al.}(2021){Wang}, {Yang}, {Fan}, {Hennawi}, {Barth},
  {Banados}, {Bian}, {Boutsia}, {Connor}, {Davies}, {Decarli}, {Eilers},
  {Farina}, {Green}, {Jiang}, {Li}, {Mazzucchelli}, {Nanni}, {Schindler},
  {Venemans}, {Walter}, {Wu}, \& {Yue}}]{wang21}
{Wang}, F., {Yang}, J., {Fan}, X., {et~al.} 2021, \apjl, 907, L1

\bibitem[{{Weinberger} {et~al.}(2018){Weinberger}, {Springel}, {Pakmor},
  {Nelson}, {Genel}, {Pillepich}, {Vogelsberger}, {Marinacci}, {Naiman},
  {Torrey}, \& {Hernquist}}]{Weinberger18}
{Weinberger}, R., {Springel}, V., {Pakmor}, R., {et~al.} 2018, \mnras, 479,
  4056

\bibitem[{{Weisskopf} {et~al.}(2007){Weisskopf}, {Wu}, {Trimble}, {O'Dell},
  {Elsner}, {Zavlin}, \& {Kouveliotou}}]{weisskopf07}
{Weisskopf}, M.~C., {Wu}, K., {Trimble}, V., {et~al.} 2007, \apj, 657, 1026

\bibitem[{{Willott} {et~al.}(2010){Willott}, {Albert}, {Arzoumanian},
  {Bergeron}, {Crampton}, {Delorme}, {Hutchings}, {Omont}, {Reyl{\'e}}, \&
  {Schade}}]{willott10}
{Willott}, C.~J., {Albert}, L., {Arzoumanian}, D., {et~al.} 2010, \aj, 140, 546

\bibitem[{{Willott} {et~al.}(2007){Willott}, {Delorme}, {Omont}, {Bergeron},
  {Delfosse}, {Forveille}, {Albert}, {Reyl{\'e}}, {Hill}, {Gully-Santiago},
  {Vinten}, {Crampton}, {Hutchings}, {Schade}, {Simard}, {Sawicki}, {Beelen},
  \& {Cox}}]{willott07}
{Willott}, C.~J., {Delorme}, P., {Omont}, A., {et~al.} 2007, \aj, 134, 2435

\bibitem[{{Willott} {et~al.}(2009){Willott}, {Delorme}, {Reyl{\'e}}, {Albert},
  {Bergeron}, {Crampton}, {Delfosse}, {Forveille}, {Hutchings}, {McLure},
  {Omont}, \& {Schade}}]{willott09}
{Willott}, C.~J., {Delorme}, P., {Reyl{\'e}}, C., {et~al.} 2009, \aj, 137, 3541

\bibitem[{{Wolf} {et~al.}(2021){Wolf}, {Nandra}, {Salvato}, {Liu}, {Buchner},
  {Brusa}, {Hoang}, {Moss}, {Arcodia}, {Br{\"u}ggen}, {Comparat}, {de
  Gasperin}, {Georgakakis}, {Hotan}, {Lamer}, {Merloni}, {Rau}, {Rottgering},
  {Shimwell}, {Urrutia}, {Whiting}, \& {Williams}}]{wolf21}
{Wolf}, J., {Nandra}, K., {Salvato}, M., {et~al.} 2021, \aap, 647, A5

\bibitem[{Wu {et~al.}(2015)Wu, Wang, Fan, Yi, Zuo, Bian, Jiang, McGreer, Wang,
  Yang, Yang, Thompson, \& Beletsky}]{wu15}
Wu, X.-B., Wang, F., Fan, X., {et~al.} 2015, Nature, 518, 512, arXiv:
  1502.07418

\bibitem[{{Yang} {et~al.}(2020){Yang}, {Boquien}, {Buat}, {Burgarella},
  {Ciesla}, {Duras}, {Stalevski}, {Brandt}, \& {Papovich}}]{yang20}
{Yang}, G., {Boquien}, M., {Buat}, V., {et~al.} 2020, \mnras, 491, 740

\bibitem[{{Yang} {et~al.}(2022){Yang}, {Fan}, {Wang}, {Lanzuisi}, {Nanni},
  {Cappi}, {Chartas}, {Dadina}, {Decarli}, {Jin}, {Keeton}, {Venemans},
  {Walter}, {Wang}, {Wu}, {Yue}, \& {Zabludoff}}]{yang22}
{Yang}, J., {Fan}, X., {Wang}, F., {et~al.} 2022, \apjl, 924, L25

\bibitem[{{Yang} {et~al.}(2021){Yang}, {Wang}, {Fan}, {Barth}, {Hennawi},
  {Nanni}, {Bian}, {Davies}, {Farina}, {Schindler}, {Ba{\~n}ados}, {Decarli},
  {Eilers}, {Green}, {Guo}, {Jiang}, {Li}, {Venemans}, {Walter}, {Wu}, \&
  {Yue}}]{yang21}
{Yang}, J., {Wang}, F., {Fan}, X., {et~al.} 2021, \apj, 923, 262

\bibitem[{{Zhou} \& {Zhang}(2010)}]{zhou10}
{Zhou}, X.-L. \& {Zhang}, S.-N. 2010, \apjl, 713, L11

\end{thebibliography}

\begin{appendix}

\section{Measuring the size of the proximity zone with an optical spectrum}
\label{sec:opt_sec}

Ly$\alpha$ proximity zones are ionised regions along the line of sight that are transparent to the quasar flux bluewards of the Ly$\alpha$ line. The surrounding inter-galactic medium (IGM) is thought to have been ionised by the UV radiation emitted by the quasar at the centre. Considering a discrete ionised H{\sc ii} region expanding in a neutral and uniform IGM, \citet{haiman01} related the radius of the quasar proximity zone (or Stromgren spheres) to the emission rate of ionising photons, $\dot{N}_{ion}$, the mean neutral hydrogen density in the IGM, $n_{\rm H\textsc{i}}$, and the lifetime of the quasar, $t_\mathrm{q}$:

\begin{equation}
    R_{\rm H\textsc{ii}} = \left( \frac{3\dot{N}_{\rm ion} \, t_\mathrm{q}}{4\pi n_{\rm H\textsc{i}}} \right)^{1/3}
.\end{equation}

The radius of the proximity zone of J0921+0007 has already been measured to be $R_\mathrm{p} = 3.05 \pm 0.45 \, \mathrm{pMpc}$  by \citet{ichimoto20} in the context of the SHELLQs survey (proper distance in Mpc). They used the low signal-to-noise ratio (S/N) discovery spectrum of the quasar taken with the Optical System for Imaging and low-Intermediate Resolution Integrated Spectroscopy (OSIRIS) at the 10.4 m Gran Telescopio Canarias \citep[GTC; 0.9 ks][]{matsuoka18}. With the availability of precise [C{\sc ii}] redshift measurements from \citet{yang21}, $z=6.5646\pm  0.0003$, we could improve upon this measurement. We re-observed J0921+0007 on February 28, 2022, with the LDSS3-C spectrograph mounted at the Magellan-Clay Telescope of the Las Campanas Observatory (Chile). A long-slit spectrum was obtained with the VPH-Red grism, which covers a range of about 6000-10500 \AA\ with a dispersion of about 1.16 \AA/px. A 1$''$ slit was used, which allowed a spectral resolution of about 4.7 \AA\ to be reached. An exposure time of 3 $\times 1200 \, \mathrm{s}$  was applied to effectively remove cosmic rays. The seeing was around 0.6-0.7$''$. We reduced the spectra with IRAF following the classic procedure of overscan subtraction, flat-field correction, and wavelength calibration. The standard star LTT3864 was observed with the same aperture slit to perform the flux calibration. Finally, the three exposures were sky-subtracted and averaged. 

The radius of the proximity zone was measured following the methodology of \cite{fan06}. The spectrum was normalised by a model for the continuum and smoothed by convolving a boxcar function of size $20 \, \AA$ with the signal. The edge of the proximity zone was then set to be the wavelength at which the continuum-normalised flux bluewards of Lyman $\alpha$ first drops below $10\%$ of the extrapolated model. In practice, the wavelength $\lambda_{\rm edge}$ is found as the wavelength at which the first of three consecutive pixels of the smoothed spectrum are below this threshold \citep{eilers17}.

The continuum bluewards of the Lyman $\alpha$ line is strongly affected by absorption and needs to be reconstructed. This can be achieved by performing a principal component analysis on the continua of lower-redshift quasars \citep{francis92,suzuki05,paris11}. Following this method, quasar spectra $q_{\rm mod}$ were modelled as the sum of an average spectrum and projections along principal components as
 
 \begin{equation}
     q_{\rm mod}(\lambda) \sim \mu (\lambda) + \sum_{j=1}^{m} c_j \xi_j(\lambda)
 ,\end{equation}where $\mu$ is an average quasar spectrum , $\xi_j$ is the j-th principal component and $c_j$ a weight specific to this quasar. In order to reconstruct the blue side of the spectrum ($\lambda < 1216 \AA$) from the red side ($\lambda > 1216 \AA$) for high-redshift quasars, two sets of principal components and associated weights were derived from lower-redshift training samples, one for the full probed wavelength range (e.g. $1020 \, \AA -2000 \, \AA$) and one only for the red range. Projections from red-range weights to full-range weights were then derived.
 
While \citet{eilers17} use a mean quasar spectrum and principal component projections derived by \citet{paris11}, \citet{ichimoto20} used results by \citet{suzuki05}, since their principal components are derived from fainter quasars, which better represent the SHELLQs quasars. We refer to these work for more details on the continuum reconstruction technique. For our quasar spectrum, we measured the proximity zone using projections from \citet{paris11}. 
 The radius of the proximity zone is then: $R_\mathrm{p}= (d_\mathrm{q} - d_{\rm edge})/(1+ z_\mathrm{q})$, where $z_\mathrm{q}$ is the quasar redshift, $d_\mathrm{q}$ and $d_{\rm edge}$ are the comoving distances derived from $z_\mathrm{q}$ and the redshift of the edge of the proximity zone: $z_{\rm edge}=\lambda_{\rm edge}/1215.67 \AA -1 $. 
 Accounting solely for the uncertainty in the [C{\sc ii}] redshift reported by \citet{yang21}, we obtained $R_\mathrm{p}= 3.05 \pm 0.01 \, \mathrm{Mpc}$, a result consistent with the measurement of \citet{ichimoto20}.
 In addition, we can account for systematic uncertainties on the [C{\sc ii}] redshift by applying a conservative offset $\delta_v = 100 \, \rm km \, s^{-1}$ in quadrature \citep{eilers20}. At $z=6.56$, such a velocity offset results in an additional redshift systematic of $\sigma_{z,\mathrm{syst}}= 0.0025$. We obtained $R_\mathrm{p}=3.05 \pm 0.13 \, \mathrm{pMpc}$. 

Since the size of the proximity zone is expected to depend on the quasar luminosity, tracing the redshift evolution of $R_p$ usually requires its measurements to be corrected to a common scale, that is, normalised to the same absolute magnitude at 1450 $\rm \AA$, $M_ {1450}$. Following the relation by \citet{eilers17}, 
 \begin{equation}
     R_{\rm p,corr}=R_{\rm p} \times 10^{0.4(27+M_ {1450})/2.35},
 \end{equation}
we obtained $R_{\rm p,corr} = 5.48 \pm 0.72 \mathrm{pMpc}$. Here we have used $M_ {1450}=-25.55 \pm 0.23$, as derived from the NIR spectral slope (see Sect. \ref{sec:nir_sec})\footnote{Assuming a steeper power-law slope of $\alpha = -1.5$, \citet{ichimoto20} adopted a brighter value $M_{1450}=-26.16 \pm 0.29$.}. The evolution of the size of quasar proximity zones with redshift has been extensively investigated \citep[e.g.][]{carilli10, venemans15,mazzuchelli17,eilers17,davies20}. From a sample of $z<6.6$ quasars, \citet{eilers17} recover a relatively shallow redshift evolution: 
 
 \begin{equation}
     R_{\rm p,corr} \approx 4.86 \,  \mathrm{pMpc} \times \left( \frac{1+z}{7}\right) ^{-1.44}
 .\end{equation}
 At $z=6.56$, the average luminosity-corrected proximity zone radius is $R_{\rm p,corr} \approx 4.34 \mathrm{pMpc} $. Similarly, using the luminosity scaling of \citet{ichimoto20}, 
 
 \begin{equation}
     R_{\rm p,corr,-25}=R_{\rm p} \times 10^{0.4(25+M_ {1450})/1.80}, 
 \end{equation}
we obtained a corrected proximity zone radius $ R_{\rm p,corr,-25}= 2.36 \pm 0.40 \, \mathrm{pMpc}$, which is larger than their prediction of the best-fit $ R_{\rm p,corr,-25} - z$ relation:  $ \bar{R_{\rm p,corr,-25}}=1.37 \pm 0.23 \, \mathrm{pMpc}$.

\begin{figure}[]
\includegraphics[width=.5\textwidth]{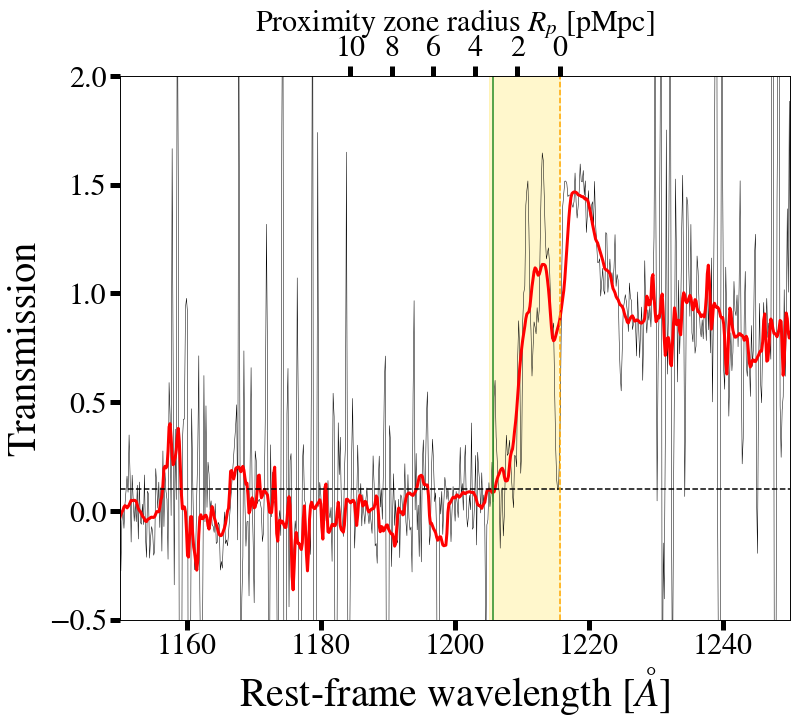}

\caption{Continuum-normalised LDSS3 spectrum of J0921+0007. The size of the proximity zone is defined here as the proper distance between the rest-frame wavelength of $Ly\alpha$ (dashed orange line) and the wavelength where the smoothed continuum normalised flux (solid red line) first drops below 0.1\% of its extrapolated value, marked here by a solid green line.  The unsmoothed continuum-normalised spectrum is shown in black.}
\label{fig:optspec}
\end{figure}

 The luminosity-scaled proximity zone radius of J0921+0007 is thus relatively large with respect to the bulk of the high-redshift quasar population. Indeed, as can be seen from the unsmoothed spectrum presented in Fig. \ref{fig:optspec}, the profile of the strong Lyman $\alpha$ line appears double-peaked with a sharp absorption feature at the exact wavelength of Lyman $\alpha$. The strong transmission bluewards of the absorption edge indicates that the quasar is embedded in a large and completely ionised region of the IGM. The unabsorbed blue wing of the Lyman $\alpha$ drives the overall line luminosity reported by \citet{matsuoka18}.

Uncertainty on the absorbed quasar continuum can significantly affect the measurement of the proximity zone. Performing Monte Carlo simulations, \citet{ichimoto20} show that the low S/N causes an uncertainty (0.71 pMpc) in their $R_\mathrm{p}$ measurement that is larger than the uncertainty due to the redshift error. We estimated the uncertainty in $R_\mathrm{p}$ imprinted by the uncertainty on the flux measurement in the red part of the spectrum. For this, we generated $1000$ spectra by perturbing the red part of the LDSS3 spectrum, at wavelengths in the observed frame at $<1216\AA \times (1+z)$. We applied Gaussian random noise accounting for the RMS of the spectrum at $9000 \, \AA$, $\sigma=2.285\times10^{-18} \, \mathrm{erg \, s^{-1}cm^{-2}}$. This is a conservative estimate of the spectral noise since the higher transmission close to the Lyman break is expected to yield lower flux uncertainties. For each of these simulated spectra, we computed the radius of the proximity zone and obtained $R_\mathrm{p}=(3.00^{+0.49}_{-0.58}) \, \mathrm{pMpc}$. 
The error related to the uncertainties bluewards of the Lyman $\alpha$ line therefore dominates uncertainties due to the redshift estimate. We stress the necessity of a higher S/N spectrum to confirm the large size of the proximity zone. We note that the LDSS3 spectrum shows no evolution in the spectral shape of Ly$\alpha$ with respect to the OSIRIS discovery spectrum presented by \citet{matsuoka18}. We repeated the measurement of the proximity zone radius of J0921+0007, using the OSIRIS discovery spectrum, accounting for the flux uncertainties redwards of the Ly$\alpha$ emission line. We obtained $R_\mathrm{p}=3.63^{+1.00}_{-0.44} \, \mathrm{pMpc}$. While this result shows larger uncertainties, it is consistent within 1$\sigma$ with the measurement on our new LDSS3 spectrum. 

The relatively large proximity zone can have two origins. The first hypothesis is that the active phase of the quasar exceeds the typical lifetime of quasars $t_\mathrm{q} \sim 10^6 \rm yr$ \citep[e.g.][]{khrykin16,davies19,khrykin21,eilers21}. This interpretation is inconsistent with the idea that the low black hole mass of J0921+0007 is indicative of a young SMBH. The luminosity-scaled proximity zone radius $ R_{\rm p,corr,-25}=2.36 \pm 0.40 \, \mathrm{pMpc}$ is relatively large for the black hole mass measured in this work \citep[][see their Fig. 8]{ichimoto20}. Alternatively the quasar has alternated between highly luminous phases and quiescent phases. The proximity zone could have grown during extremely luminous phases of the AGN. Such large amplitude variability in the ionising continuum emission is indeed expected in NLS1s \citep{collier01,romano02}. The SED of NLS1s is also potentially quite different to that of broad-line quasars, which can further affect the ionisation balance in the proximity zone. 
J0921+0007 is the second most Lyman $\alpha$ luminous quasar of the SHELLQs survey \citep{matsuoka16,matsuoka18,matsuoka19,matsuoka22}, which implies that the broad-line region gas is exposed to strong ionising radiation. This is also confirmed by its unusually high X-ray luminosity (Fig. \ref{fig:lx_plot} and \ref{fig:aox}). We therefore propose that the quasar is currently undergoing a super-Eddington accretion phase that generates powerful UV/X-ray radiation, which in turn ionises the surrounding IGM efficiently. It is in such phases that the proximity zone grows to relatively large radii. The ionising continuum radiation of NLS1s is, however, expected to show large amplitude short-term variability. The response of non-equilibrium blinking light-bulb quasar models has been studied by \citet{davies20}; however, the timescales on which NLS1s are expected to show X-ray/UV variability (hours to days) are much smaller than the shortest timescales explored in this work (100 yr). Capturing X-ray variability on timescales of days for this quasar would require a long monitoring campaign \citep[>180 ks; e.g.][]{nanni18,yang22}. The confirmation of short timescale X-ray variability could explain the relatively large proximity zone of J0921+0007. This could imply phases of even more extreme X-ray loudness.
In addition, a long exposure would put more stringent constraints on the spectral shape of the source.

\end{appendix}

\end{document}